\documentclass[%
reprint,
superscriptaddress,
nofootinbib,
amsmath,amssymb,
aps,
prl,
]{revtex4-1}
\usepackage{xr-hyper}
\usepackage{hyperref}
\usepackage{graphicx} 
\usepackage{dcolumn}
\usepackage{bm}
\usepackage{braket}
\usepackage{dsfont}
\usepackage{color,colortbl}
\usepackage{chemmacros}
\DeclareUnicodeCharacter{2212}{-}
\usepackage{multirow}
\usepackage{subcaption}
\usepackage{mwe}
\usepackage{booktabs}
\usepackage{caption}
\usepackage{lipsum}
\usepackage{babel,blindtext}
\usepackage{amsmath}
\usepackage[toc,page]{appendix}
\usepackage[symbol*]{footmisc}
\usepackage{float}

\usepackage{natbib}
\usepackage{filecontents}

\makeatletter

\newcommand*{\addFileDependency}[1]{
\typeout{(#1)}
\@addtofilelist{#1}
\IfFileExists{#1}{}{\typeout{No file #1.}}
}\makeatother

\newcommand*{\myexternaldocument}[1]{%
\externaldocument{#1}%
\addFileDependency{#1.tex}%
\addFileDependency{#1.aux}%
}

\myexternaldocument{SI}

\begin{document}
\title{Adsorbate-Induced Surface Amorphisation of Catalytic Nanoparticles}
\title{Do facets exist on nanoparticle catalysts?}
\title{A chocolate candy model of nanoparticle catalysts}
\title{Surface plasticity in nanoparticle catalysts}
\title{Surface roughening in nanoparticle catalysts}


\author{Cameron J. Owen$^{*,\dagger}$}
\affiliation{Department of Chemistry and Chemical Biology, Harvard University, Cambridge, Massachusetts 02138, United States}
\affiliation{John A. Paulson School of Engineering and Applied Sciences, Harvard University, Cambridge, Massachusetts 02138, United States}

\author{Nicholas Marcella$^{*}$}
\affiliation{Department of Chemistry, University of Illinois, Urbana, Illinois 61801, United States}

\author{Christopher R. O'Connor$^{*}$}
\affiliation{Rowland Institute at Harvard, Harvard University, Cambridge, Massachusetts 02142, United States}

\author{\\Taek-Seung Kim}
\affiliation{Rowland Institute at Harvard, Harvard University, Cambridge, Massachusetts 02142, United States}

\author{Ryuichi Shimogawa}
\affiliation{Department of Materials Science and Chemical Engineering, Stony Brook University, Stony Brook, New York 11794, United States}
\affiliation{Mitsubishi Chemical Corporation, Science and Innovation Center, 1000 Kamoshida-cho, Aoba-ku, Yokohama 227-8502, Japan}

\author{Clare Yijia Xie}
\affiliation{John A. Paulson School of Engineering and Applied Sciences, Harvard University, Cambridge, Massachusetts 02138, United States}

\author{Ralph G. Nuzzo}
\affiliation{Department of Chemistry, University of Illinois, Urbana, Illinois 61801, United States}

\author{Anatoly I. Frenkel$^{\dagger}$}
\affiliation{Department of Materials Science and Chemical Engineering, Stony Brook University, Stony Brook, New York 11794, United States}
\affiliation{Chemistry Division, Brookhaven National Laboratory, Upton, New York 11973, United States}

\author{Christian Reece$^{\dagger}$}
\affiliation{Rowland Institute at Harvard, Harvard University, Cambridge, Massachusetts 02142, United States}

\author{Boris Kozinsky$^{\dagger}$}
\affiliation{John A. Paulson School of Engineering and Applied Sciences, Harvard University, Cambridge, Massachusetts 02138, United States}
\affiliation{Robert Bosch LLC Research and Technology Center, Watertown, Massachusetts 02472, United States}

\def\thefootnote{$*$}\footnotetext{These authors contributed equally.}\def\thefootnote{\arabic{footnote}}

\def\thefootnote{$\dagger$}\footnotetext{Corresponding authors\\C.J.O., E-mail: \url{cowen@g.harvard.edu}\\A.I.F., E-mail: \url{anatoly.frenkel@stonybrook.edu}\\C.R., E-mail: \url{christianreece@fas.harvard.edu}\\B.K., E-mail: \url{bkoz@g.harvard.edu}}\def\thefootnote{\arabic{footnote}}

\begin{abstract}
Supported metal nanoparticle catalysts are vital for the sustainable production of chemicals, but their design and implementation are limited by the ability to identify and characterise their structures and atomic sites that are correlated with high catalytic activity \cite{Snitkoff-Sol_Friedman_Honig_Yurko_Kozhushner_Zachman_Zelenay_Bond_Elbaz_2022,Vogt_Weckhuysen_2022}. Identification of these ``active sites'' has relied heavily on extrapolation to supported nanoparticle systems from investigation of idealized (faceted) surfaces, experimentally using single crystals \cite{Somorjai_1985} or supported nanoparticles \cite{frenkel_view_2001,mostafa2010shape} which are always modelled computationally using slab or regular polyhedra models  \cite{Nørskov_Bligaard_Rossmeisl_Christensen_2009}. However, the ability of these methods to predict catalytic activity remains qualitative at best \cite{Beck_Paunovic_VanBokhoven_2023,Bruix_Margraf_Andersen_Reuter_2019}, as the structure of metal nanoparticles in reactive environments has only been speculated from indirect experimental observations, or otherwise remains wholly unknown. Several dynamic investigations, e.g. by density functional theory (DFT), have demonstrated some progress towards understanding these catalytically active sites through an atomistic lens, but these methods are inherently hindered in both time- and length-scales \cite{PhysRevB.78.121404,vila2017anomalous}, meaning that they also require extrapolation to explain experimental results. Here, by circumventing these limitataions for highly accurate simulation methods, we provide direct atomistic insight into the dynamic restructuring of metal nanoparticles by combining in situ spectroscopy with molecular dynamics simulations powered by a machine learned force field. We find that in reactive environments, nanoparticle surfaces evolve to a state with poorly defined atomic order, while the core of the nanoparticle may remain bulk-like. These insights prove that long-standing conceptual models based on idealized faceting for small metal nanoparticle systems are not representative of real systems under exposure to reactive environments. We show that the resultant structure can be elucidated by combining advanced spectroscopy and computational tools. This discovery exemplifies that to enable faithful quantitiative predictions of catalyst function and stability, we must move beyond idealized-facet experimental and theoretical models and instead employ systems which include realistic surface structures that respond to relevant physical and chemical conditions.
\end{abstract}

\maketitle

Catalysis by design has largely been driven by theoretical and experimental screening using idealized (faceted) slab models, from which various properties (e.g., structure, kinetics) are evaluated, and catalytic activity can be predicted \cite{Liu_Shih_Deng_Ojha_Chen_Luo_McCrum_Koper_Greeley_Zeng_2024,Reece_Redekop_Karakalos_Friend_Madix_2018,Seh_Kibsgaard_Dickens_Chorkendorff_Nørskov_Jaramillo_2017,Marcella2022}. While these tools have proven to be extremely powerful for well defined systems, they are only, at best, able to generate qualitative predictions \cite{Mou_Pillai_Wang_Wan_Han_Schweitzer_Che_Xin_2023}. Problematic within these bounds is the fact that planar single crystal and supported metal nanoparticle systems have long been known to be structurally dynamic \cite{Somorjai_1992}, which has large implications on catalytic activity \cite{Somorjai_Carrazza_1986}. Even in the absence of a reaction environment, surface distortion of nanocrystals (3-5~nm) occur and can be distinct from single crystal materials \cite{huang2008coordination}. Further complicating the appliciability of single crystal materials to model supported nanoparticles is the dynamic disruption of single crystalline surfaces to form nanoparticles under reaction environments through the mass transport of surface atoms driven by adsorbates \cite{Xu_Papanikolaou_Lechner_Je_Somorjai_Salmeron_Mavrikakis_2023,Eren_Zherebetskyy_Patera_Wu_Bluhm_Africh_Wang_Somorjai_Salmeron_2016, Eren_Torres_Karslıoğlu_Liu_Wu_Stacchiola_Bluhm_Somorjai_Salmeron_2018, Roiaz_Falivene_Rameshan_Cavallo_Kozlov_Rupprechter_2019,Tao_Dag_Wang_Liu_Butcher_Bluhm_Salmeron_Somorjai_2010}. For the past two decades, the effect of reactive environments on surface atoms has been inferred for nanoparticles ranging from very small (1~nm) \cite{li_noncrystalline--crystalline_2013, vila2017anomalous, Vila2008} to larger (3~nm) \cite{frenkel_view_2001, Erickson2014} diameters using extended X-ray absorption fine structure spectroscopy (EXAFS), as indicated by variations in bond length distribution. In the best cases, EXAFS is used in bespoke correlative studies to build models of surface-level mechanisms, such as EXAFS in combination with scanning transmission electron microscopy (STEM) and theory \cite{Li2021}, but this is only valid for very small nanoparticle systems where the surface atoms represent a substantial fraction of the total number of atoms. It is thus impossible to resolve the exact atomistic details of structural evolution. Atomistic modeling of these effects also remained intractable due to the complexity and computational cost of the necessary first principles molecular dynamics (MD) simulations. Now, it has become possible to couple reactive MD simulations, powered by machine learned force fields (MLFFs) \cite{Vandermause2020,Vandermause2022,Batzner2021E3-EquivariantPotentials,Musaelian2023,NEURIPS2022_4a36c3c5,Shapeev2016MomentPotentials}, with correlative in situ infrared and X-ray absorption spectroscopy to greatly enhance the atomic and temporal resolution of such analysis schemes. This combination now makes it feasible to gain fully atomistic understanding into the dynamic structural responses of catalytically-relevant small metal nanoparticle systems \cite{owen2023unraveling} and examine their function and mechanisms.

\begin{figure*}
\centering
\includegraphics[width=\textwidth]{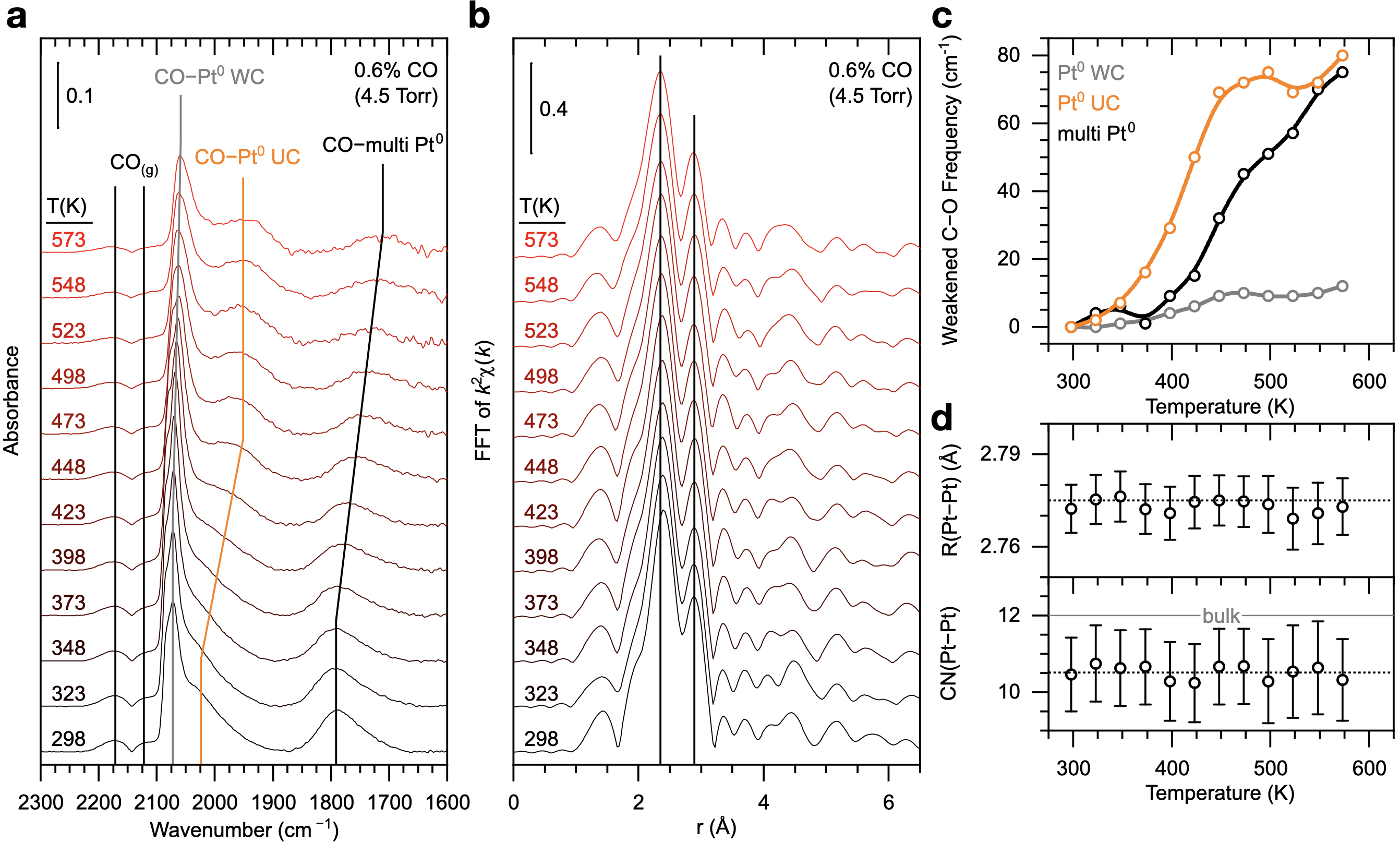}
\caption*{\textbf{Fig. 1 \textbar \space DRIFTS and EXAFS spectra of the 2~nm Pt/SiO$_2$ catalysts in 4.5 Torr of 0.6 \% CO balanced in Ar.} \textbf{a}, Temperature dependent DRIFTS spectra of CO adsorbed on Pt/SiO$_2$ with the frequency positions of the three main features highlighted. \textbf{b}, The magnitude of the Fourier transformed k2-weighted EXAFS (FT-EXAFS) spectra as a function of temperature. \textbf{c}, Frequency shift of the three main DRIFTS features relative to the peak positions of the initial spectra at 298 K. \textbf{d}, The temperature dependent interatomic distance (R) and Pt-Pt coordination number (CN) calculated using EXAFS curve fitting.}
\label{fig:drifts}
\end{figure*}

\subsection*{Nanoparticle surfaces dramatically respond to CO}
Here we report the role of temperature and gaseous exposure on the loss of atomic order on nanoparticle surfaces using a combination of experimental characterization and atomistic simulation techniques. The dynamic structural response of a 2~nm Pt nanoparticle catalyst supported on SiO$_2$ (2~nm Pt/SiO$_2$) to a gas phase of carbon monoxide (CO) was evaluated using in situ Diffuse Reflectance Infrared Fourier Transform spectroscopy (DRIFTS) for study of the surface structure and X-ray Adsorption Spectroscopy (XAS) for study of the bulk structure. As demonstrated in Fig. 1a, the DRIFT spectra are sensitive to surface-level reconstructions, as the C-O stretching frequency is an effective measure of structural changes to metal surfaces since the adsorption couples with donation of electron density to or from the CO molecule which shifts the vibrational frequency \cite{Hollins_1992}. Three features are observed (Fig. 1a) corresponding to a CO molecule adsorbed linearly on a well-coordinated (e.g. (111) or (100) terraces) site (CO-Pt$^0$ WC), adsorbed linearly on under-coordinated (e.g. edge or step-like) sites (CO-Pt$^0$ UC), and adsorbed between multiple Pt atoms (CO-multi Pt$^0$) \cite{Kale_Christopher_2016,kim_interrogating_2024}. As the 2~nm Pt/SiO$_2$ catalyst was heated from 298 to 573 K, dramatic shifts in the vibrational frequencies of the adsorbed CO-Pt$^0$ UC and CO-multi Pt$^0$ to lower frequencies were observed which indicates a weaker C-O bond (Fig. 1b). This shift is much larger in magnitude than would be expected for changes in the strength of CO binding with coverage due to repulsive CO dipole-dipole coupling as observed for CO-Pt$^0$ WC \cite{Hollins_1992,crossley1980adsorbate}, which suggests that a dramatic surface structural change occurs that changes the bonding between surface Pt and CO. 

\begin{figure*}
\centering
\includegraphics[width=\textwidth]{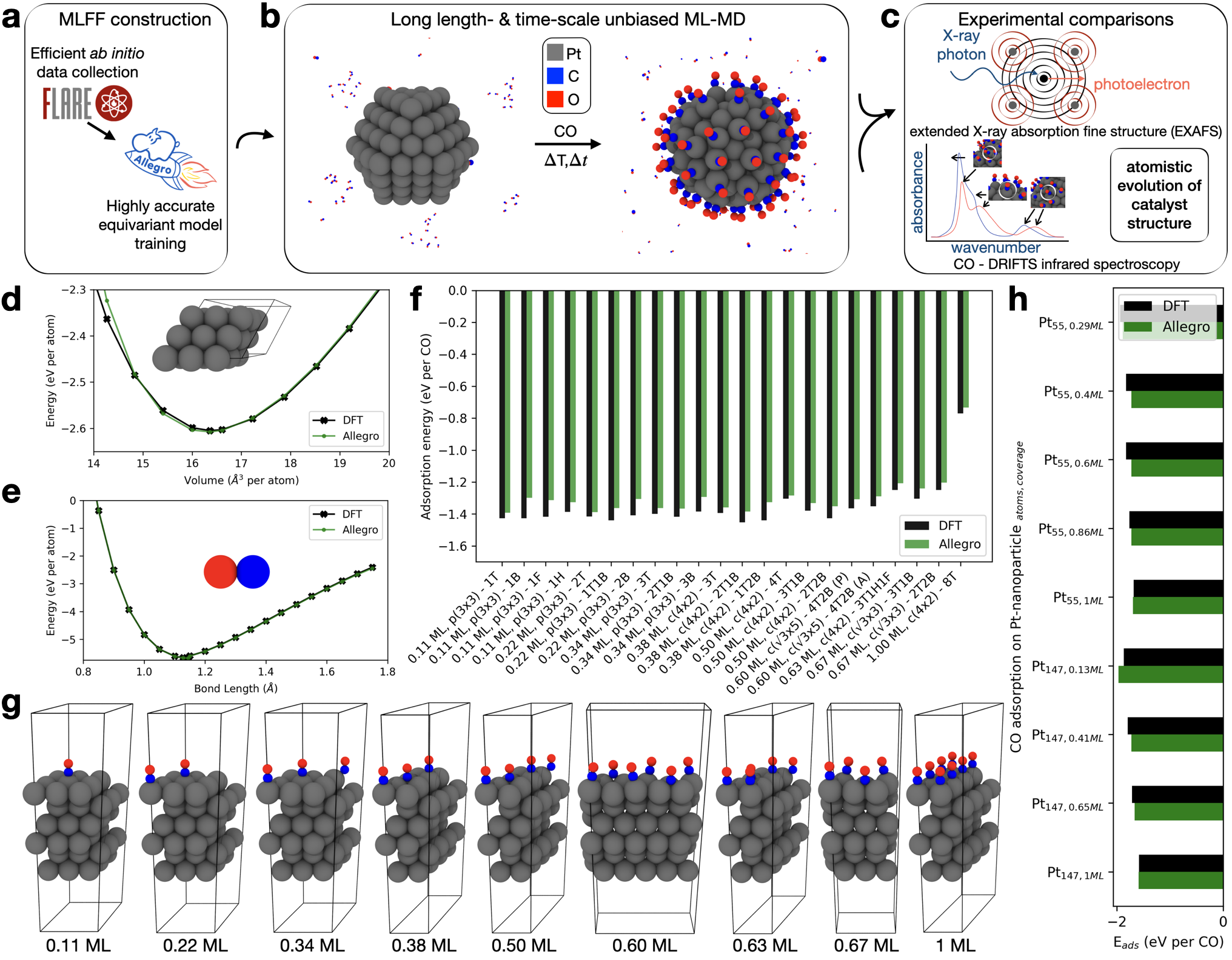}
\caption*{\textbf{Fig. 2 \textbar \space
Construction pipeline to yield an Allegro MLFF for atomistic understanding of CO/Pt systems.} \textbf{a}, Efficient \textit{ab initio} frame collection is completed using FLARE active-learning and sequentially paired with training an Allegro MLFF, which ultimately yields a flexible model for simulation of platinum single-crystal and NP catalysts interacting with CO. \textbf{b}, The MLFF is then used to simulate a comprehensive suite of NP and single-crystal systems, where both the input morphology and environment can be varied in unbiased ML-MD simulations to understand preferential evolution of the atomistic structure. \textbf{c}, Unbiased simulation results can then be directly compared to a combination of experimental characterization techniques, such as EXAFS and CO-DRIFTS used here, to (1) provide confidence in the simulated structural evolutions \textit{in silico} and (2) provide direct atomistic resolution to explain the complicated experimental data from DRIFTS and EXAFS. \textbf{d}, Bulk equation of state (energy (eV per atom) versus volume (\AA{}$^3$ per atom) predicted by Allegro (green) and DFT (black). \textbf{e}, Total energy (eV per atom) of CO under bond compression and tension predicted by DFT and Allegro. \textbf{f}, CO adsorption energy (eV per CO) on Pt(111) as a function of site preference and coverage predicted by DFT and Allegro. \textbf{g}, Snapshots of the minimum energy structures for CO adsorption on Pt(111) from \textbf{f}. \textbf{h}, CO adsorption energy (eV per CO) on cuboctahedral Pt NPs across sizes and CO coverages by DFT and Allegro.}
\label{fig:toc}
\end{figure*}

\begin{figure*}
\centering
\includegraphics[width=1\textwidth]{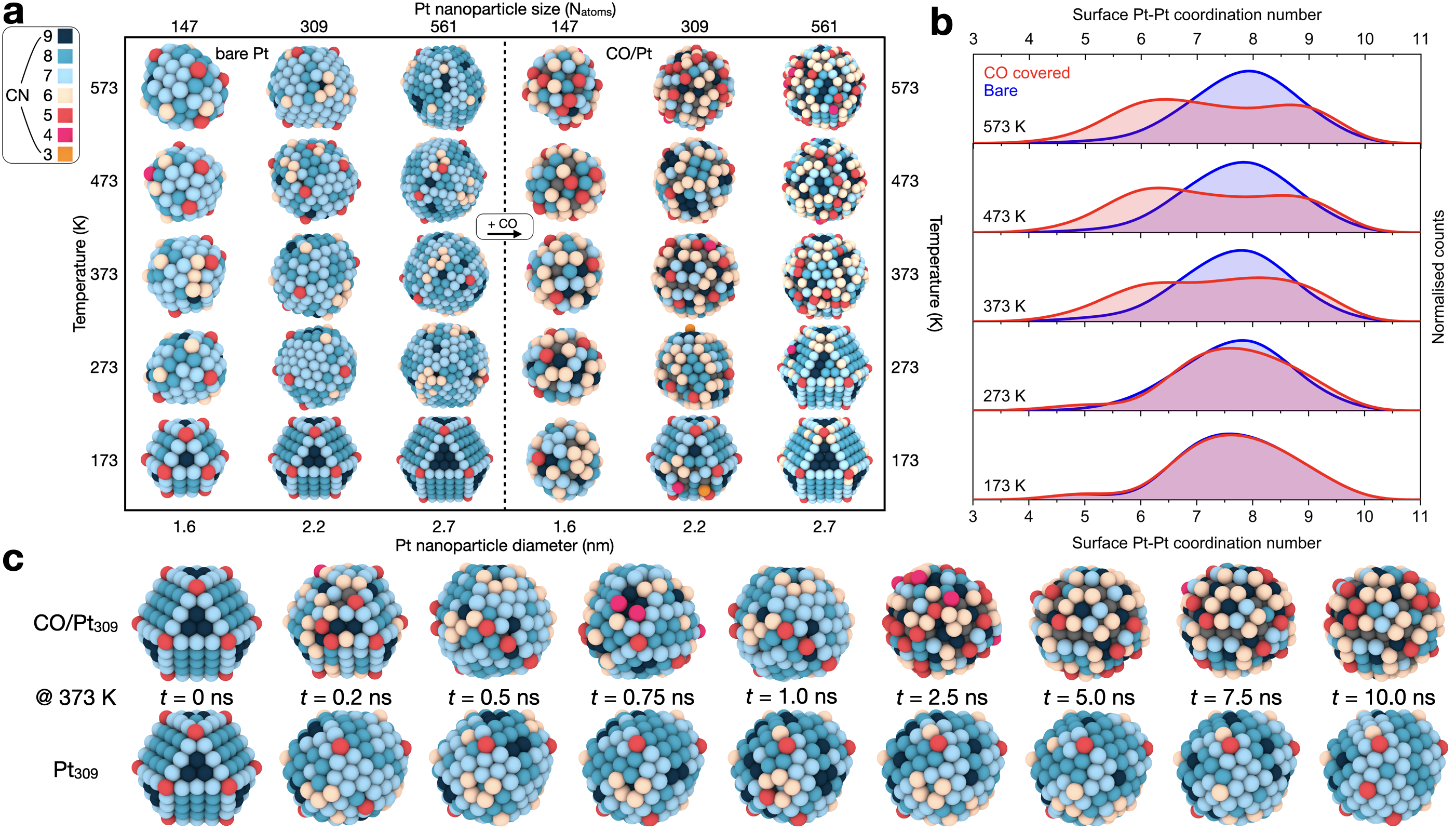}
\caption*{\textbf{Fig. 3 \textbar \space Nanoparticle evolutions with and without CO using the MLFF.} \textbf{a} Simulation snapshots taken at the end of each trajectory for Pt$_{147}$, Pt$_{309}$, Pt$_{561}$, and the CO analogues of each. CO is removed for clarity and the surface atoms are colored according to their Pt-Pt coordination. \textbf{b} Time series examples for a CO-exposed and bare Pt$_{309}$ particle at 373 K. \textbf{c} Coordination analysis for the Pt$_{561}$ atom systems in \textbf{a}, denoting the shift from bare (blue) to CO-exposed (red).
}
\label{fig:copt}
\end{figure*}

\subsection*{Nanoparticle cores do not dramatically respond to CO}
Analogous in situ XAS measurements were performed on the same Pt/SiO$_2$ catalyst to rationalize if the dramatic shifts in the DRIFTS spectra were related to a bulk restructuring of the nanoparticle. The Fourier transform (FT) magnitude of the k$^2$-weighted EXAFS spectrum (Fig. 1c) remains unchanged, and only shows a marginal decrease in intensity with temperature, which arises due to temperature dependent bond length disorder \cite{Frenkel_Hills_Nuzzo_2001}. Using EXAFS fitting, the coordination number (CN), interatomic distance (R), and Debye-Waller factor ($\sigma^2$) were quantitatively assessed. No significant restructuring of the 2~nm Pt nanoparticles was observed in response to the CO environment as evidenced by a consistent interatomic distance of 2.773 $\pm$ 0.003 \AA{} and a Pt-Pt coordination number of 10.5 $\pm$ 0.3 (Fig. 1d), while the Debye-Waller factor was found to increase linearly with temperature consistent with the correlated Einstein model (Supplementary Fig. 1) \cite{FrenkelRehr1993}. The temperature-dependent structural changes of 2~nm Pt/SiO$_2$ in CO by surface-sensitive DRIFTS measurements but absence in more bulk-sensitive XAS measurements can be rationalized by surface restructuring that does not extend into the bulk. A local, atomistic understanding of the entire nanoparticle structure is thus required to resolve these apparently distinct phenomena of the catalyst surface and bulk structure, which necessitates employing a MLFF to simulate the temperature-dependent response of various nanoparticles $\approx$ 2~nm in diameter, in the presence and absence of CO. 

\subsection*{Machine learned dynamics deconvolve these disparate signals}
The workflow to develop the MLFF is described in Fig. 2a. We start with Bayesian active learning to efficiently construct a set of \textit{ab initio} data, from which the final MLFF is trained and used to simulate the Pt nanoparticle and single crystal systems. The active learning algorithm, as implemented in the FLARE framework \cite{Vandermause2022}, constructs a surrogate energy model and propagates the atomistc dynamics, automatically invoking and learning from expensive \textit{ab initio} computations whenever relevant and unfamiliar (high predictive uncertainty) atomic evironments are encountered. This approach has been successfully employed to investigate materials systems inaccesible to traditional computational approaches \cite{Owen2024_au,owen2023unraveling,owen2024unbiased,Vandermause2022,Johansson2022Micron-scaleLearning,Xie2023Uncertainty-awareSiC,Xie2021BayesianStanene}. Using the set of structures generated by active learning, we train the MLFF production model using the Allegro architecture. This deep equivariant neural network architecture has been proven to exhibit the leading combination of accuracy, stability and scalability across a wide range of materials systems\cite{Musaelian2023,10.1145/3581784.3627041}. Once trained, the MLFF was evaluated against DFT to determine it's efficacy in predicting low energy bulk (Fig. 2d), gas-phase (Fig. 2e), adsorption (Fig. 2f,g), and nanoparticle structures (Fig. 2h). Given the success of the MLFF to provide DFT-level accuracy calculations at a fraction of the cost, as demonstrated in Fig. 2, the MLFF was then employed to perform reactive ML-accelerated MD simulations (ML-MD). We explicitly examine the real-time response of nanoparticle systems to gaseous species, allowing for in silico evolution and determination of nanoparticle structure, generating information that is directly comparable with in situ spectroscopy measurements, without any prior assumptions. This approach has significant benefits over other thermodynamic techniques (e.g. basin hopping algorithms \cite{doi:10.1021/jacs.2c10179,D1SC03827C,doi:10.1021/acscatal.0c01971}) which do not account for the kinetics of restructuring and do not explicitly capture the actual time scale and mechanisms of the dynamic interactions between the adsorbate and the surface. Specifically, to match the experimental nanoparticle distribution of 2.0 $\pm$ 0.4 nm (Supplementary Fig. 2) and reaction conditions, free-standing Pt nanoparticles consisting of 147, 309, and 561 atoms of $\approx$ diameter of 1.6, 2.2, and 2.7 nm were simulated both in the absence of CO and in the presence of 50-750 Torr CO in the temperature range from 148 to 623 K (Fig. 3). The difference in pressure was employed to accelerate the time-scale by which the nanoparticles reached equilibrium coverage, and was checked carefully with regards to the resulting coverage in comparison to DRIFTS measurments at higher pressures (e.g. 20\% CO/Ar in Suppl. Fig. 3). The inclusion of an explicit SiO$_2$ support was deemed to not be critically important due to the weak interaction between Pt and SiO$_2$, as evidenced by the lack of sufficient wetting of the nanoparticles to the surface \cite{Ahmadi_Timoshenko_Behafarid_Roldan_Cuenya_2019,Simonsen_Chorkendorff_Dahl_Skoglundh_Sehested_Helveg_2011}. Particles with truncated cuboctahedral morphology and faceting were explicitly created to initialize each MD trajectory, as this shape has been shown to be present for supported Pt nanoparticles on SiO$_2$ and CeO$_2$ using high-resolution transmission electron microscopy \cite{Ahmadi_Timoshenko_Behafarid_Roldan_Cuenya_2019,Simonsen_Chorkendorff_Dahl_Skoglundh_Sehested_Helveg_2011,Li2021}. This choice is not critical, as the shape is allowed to relax and evolve along all degrees of freedom during the simulation under the effects of temperature and CO pressure. 

From visualization of the structural evolution (Fig. 3a) it is clear that both temperature and CO have drastic and disparate effects on the surface morphology of the nanoparticles. Overall, there is a greater concentration of surface atoms with low coordination number in the presence of CO at elevated temperatures. In the absence of CO, the surfaces of all nanoparticles considered significantly restructure to form a closely packed smooth pseudo-(111) surface characterized by a face-centered cubic pattern with a lower coordination number than an idealized 111 facet. The complete restructuring of the distinct (100) and (111) facets of the initial truncated octahedron nanoparticle structure to uniform pseudo-(111) terraces may explain why catalytic phenomenon observed with (111) single crystal model surface experiments and computational slab models have been reasonably successful for understanding nanoparticle catalyst function. However, the difference between the pseudo-(111) surfaces on nanoparticles and idealized (111) terrace models could contribute to the inability to quantitatively predict nanoparticle catalytic behavior using simplified experimental models, similar to previous demonstrations of the effect of strain on Pt reactivity \cite{Liu_Shih_Deng_Ojha_Chen_Luo_McCrum_Koper_Greeley_Zeng_2024}. In the pseudo-(111) surface we observe the emergence of `rosette' patterns, similar to what we have previously reported on small (55 atom) nanoparticle systems with increasing temperature and hydrogen coverage in a H$_2$ environment\cite{owen2023unraveling}. The occurence here of `rosette' structures on larger nanoparticles is ascribed to the decreased surface-area-to-volume ratio that allows for more flexible relaxation profiles of the surface atoms due to decreased surface tension. In the presence of CO, the long-range order of the surface is entirely disrupted and we observe a large increase of the number of surface Pt atoms with lower coordination, giving the appearance of a rough surface (Figure 3b). This resembles the effect of strain on the reactivity on Pt surfaces \cite{Liu_Shih_Deng_Ojha_Chen_Luo_McCrum_Koper_Greeley_Zeng_2024}, as strain also influences the effective coordination of the atoms, directly affecting the electronic structure available for catalysis. These subtle modifications of the atomic structure corresponding to dramatic changes in electronic structure and catalytic activity highlight the necessity for simulation methods to capture non-idealized structures, which underly all of catalysis. 

The time dependent evolution of the nanoparticle surface demonstrates a rapid transformation to a smooth pseudo-(111) surface within 0.5~ns independent of the gas environment but occurs slightly slower in the presence of CO (Fig. 3c). However, in the presence of CO, the pseudo-(111) surface evolves to yield highly undercoordinated Pt atoms within 5~ns. The evolution of the pseudo-(111) surface to a disordered undercoordinated surface is attributed to the strong lateral repulsion between adsorbed CO species where a stronger anchoring Pt-C bond weakens the bonding of the Pt atom host to other Pt atoms \cite{Li2021,Wang_fragmentation_2024}. Further, the evolution towards a disordered, undercoordinated surface in the presence of CO can be attributed to the modulation of the intermetallic cohesion strengh arising from the charge transfer from CO bound in a top-site orientation to surface Pt \cite{kalhara2018co}, whereas a more ordered Pt surface was reported in an H$_2$ environment where dissociated H atoms withdraw electron density from the Pt surface \cite{catal8100450}.

\begin{figure*}
\centering
\includegraphics[width=1\textwidth]{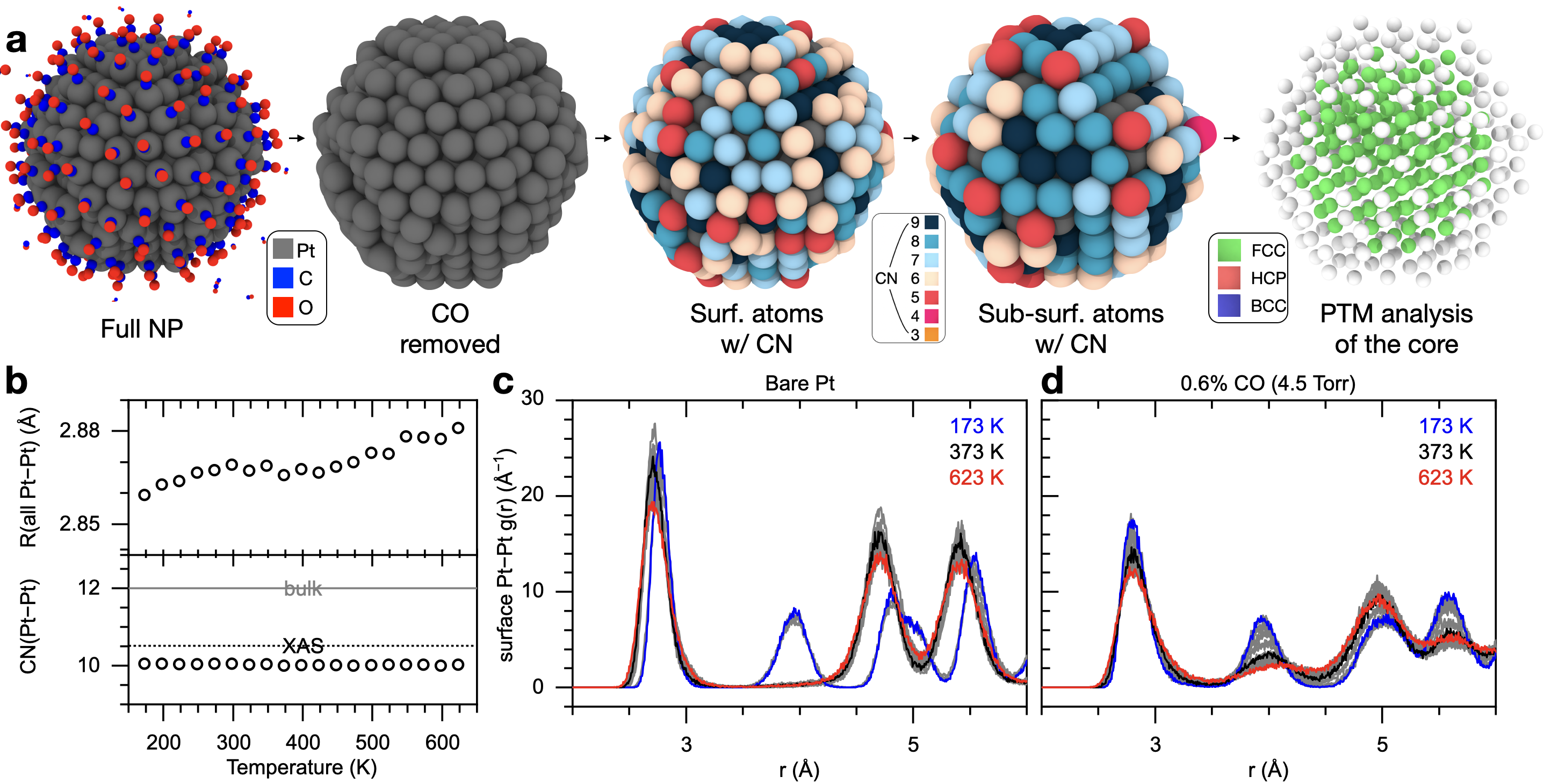}
\caption*{\textbf{Fig. 4 \textbar \space MLFF shows that CO only distorts the NP surface.} \textbf{a}, Snapshot of CO/Pt$_{561}$ evolved for $\approx$ 10 ns at 373 K with different structural modifications to isolate the surface and core environments. The CO and surface atoms are removed in order to assess the crystallinity of the NP core using Polyhedral Template Matching (PTM) in Ovito \cite{ovito}. \textbf{b}, Pt-Pt interatomic distance (R) and coordination number average across the final 1 ns of simulation time for all temperatures from 148 - 623 K. \textbf{c}, Surface Pt-Pt radial distribution function for the bare particles from low (blue) to high temperature (red) in the ML-MD simulations. \textbf{d}, Surface Pt-Pt radial distribution function for the CO exposed particles from low (blue) to high temperature (red) in the ML-MD simulations.
}
\label{fig:np}
\end{figure*}

To rationalize the seemingly contradictory spectroscopic measurements, and to support the hypothesis that the surface structure responds to the environment while the inner core does not, we use ML-MD simulations to examine the Pt surface atoms in isolation. Visualization on Fig. 4a confirms the disparity of the surface and bulk responses, thus reconciling observations of the surface level effects (which are probed by DRIFTS) and bulk effects (probed by XAS) and providing direct atomistically resolved interpretation of the entire set of measurements from a single simulation representing realistic conditions. The ML-MD results accurately reproduce the EXAFS signal, with a consistent interatomic distance of $\approx$ 2.87 \AA{} and a Pt-Pt coordination number of $\approx$ 10 (Fig. 4b) across the same temperature domain as that of the experimental measurments. The increased interatomic distance and sensitivity to temperature in the simulations arises from the underlying DFT method overestimating the bond length of Pt by 2.8 \%. To rationalize the DRIFTS results, the radial distribution function of the surface Pt atoms was calculated over the nanoparticles without and with the CO atmosphere (Fig. 4c,d, respectively). In the absence of a CO environment (Fig. 4c), the nanoparticle surface rapidly reconstructs to a crystalline pseudo-(111) phase as the temperature is increased, as denoted by the loss of the feature at 4 \AA{} and the persistence of sharp peaks at $\approx$ 4.6 and 5.4 \AA{}. When CO is present, the radial distribution function is much broader and has a higher value between well-defined peaks, corresponding to the increased amorphous character of the surface atoms. The gradual broadening of the radial distribution function with increasing temperature suggests that the transition is not discrete, but continuous. When coupled with the coordination analysis (Fig. 3a), this gradual amorphization of the nanoparticle surface with temperature aligns well with the measured DRIFTS spectra where a gradual change in vibrational frequency of adsorbed CO was observed (Fig. 1a,c). The ML-MD simulations thus indicate that the two different in situ spectroscopy measurements are not contradictory, and elucidate how different regions of the nanoparticle are responding to environmental conditions. Hence, taking the atomistic insights from the MLFF and ML-MD simulations, paired with the resolution of EXAFS and DRIFTS, we provide a comprehensive interpretation of the response of Pt nanoparticle surfaces to CO, where a disordered, undercoordinated rough layer of atoms bound to adsorbates sits atop a more bulk-like FCC lattice structure. This finding has enormous implications in the larger field of nanomaterials and particularly heterogeneous catalysis, as nanoparticle structure dictates performance \cite{Yang2023,Liu_Shih_Deng_Ojha_Chen_Luo_McCrum_Koper_Greeley_Zeng_2024}.

\subsection*{Nano-catalyst surfaces cannot be assumed to have idealized facets}
In this work we have demonstrated that the structural morphology of nanoparticle systems in reactive environments is far more complex than originally understood. Idealized faceting of nanoparticle systems, which has been the foundation of catalyst design for decades, is not reliable for the interpretation of real catalytic systems, particularly for ones under reactive environments and under increased temperature. While hints as to these responses have been available within the literature, the multimodal investigation presented here allows for in silico atomistic understanding of such phenomena, with excellent agreement to disparate experimental techniques. The rapid reconstruction of the nanoparticle surface to reveal under-coordinated and increasingly amorphous structures in the presence of CO explains why theoretical calculations have especially struggled to accurately recreate experimentally measured reaction kinetics over these nanoparticle systems \cite{doi:10.1021/acs.jpca.6b06770,doi:10.1021/jacs.2c10179,D1SC03827C,doi:10.1021/acscatal.0c01971}, especially given difficulties for the underlying quantum mechanical methods to model adsorption of CO/Pt \cite{PhysRevB.69.161401,doi:10.1021/jp002302t}. More generally, computational modeling of realistic systems can be prone to error in model selection where input geometry for ab initio molecular dynamics simulations can substantially affect the conclusions. Hence, the work presented here moves towards a large reduction in such error, as the material systems can be evolved according to the underlying physics across appropriate length- and time-scales to interpret experimental signals. To support this claim, we make evident the importance of combining multiple experimental characterization methods when considering nanoparticle structure, and how ML-MD simulations are the critical link for an unbiased view into nanoparticle structure via in silico pretreatments. It would be a mistake to assume that the phenomena reported in this work are unique to Pt or CO environments, as it has been long hypothesized, but not confirmed with atomstic resolution using such a multimodal approach, that adsorbates will restructure catalytic surfaces, and as such, this study has large knock-on effects for catalyst design. Importantly, the large deviation in coordination number of the surface atoms as a function of temperature and adsorbate exposure means that long-standing chemical intuition may not indeed be successful a priori in catalyst design, prompting the need for direct atomistic simulation to first determine the plausible distribution of active site geometries. This advancement can then be coupled with existing scaling relationships that have proven extremely valuable in recent decades, namely d-band theory \cite{Hammer1995WhyMetals} for adsorption strengths, as the ML-MD can provide the distribution of realistic coordination numbers for the active surface atoms. 

\subsection*{Methods}
\subsection*{Density Functional Theory}
All DFT calculations were carried out using the Vienna \textit{Ab initio} Simulation Package (VASP) \cite{Kresse1996}, which describes electron-ion interactions using the projector-augmented wave (PAW) formalism. We studied CO adsorption on Pt NPs, (111), (110), and (100) in FLARE active learning using periodic density functional theory with the vdW-DF exchange correlation functional and a plane-wave basis set with a cutoff kinetic energy of $500$ eV. For validation of the Allegro MLFF, the bulk Pt cell was modeled with a $3\times3$ supercell (Fig. 2d) and the Pt(111) surface was modeled with a supercell containing six-atomic layers (Fig. 2f), with a vacuum layer of $10\AA$ between periodic slabs to minimize interactions between repeated slabs. The number of CO molecules per unit cell and their adsorption sites (top, bridge, HCP, and FCC) were extensively sampled, especially for the low coverage structures. During geometry optimization, the bottom two layers of the Pt slab were constrained at the bulk positions with optimized lattice constants of 4.03 \AA{}, while the top four layers and the adsorbed CO molecules are allowed to relax. All geometries were optimized until the energy between consecutive steps changed by less than $10^{-5}$ eV. The energies for all structures calculated using DFT were also calculated with our MLFF in LAMMPS \cite{THOMPSON2022108171}. The adsorption energies per CO molecule, $n$, from DFT and MLFF were then calculated and compared (Fig. 2f,h).

\subsection*{FLARE Bayesian Active Learning}
The MLFF architecture used for the CO/Pt systems during \textit{ab initio} data collection is from the Fast Learning of Atomistic Rare Events (FLARE) open-source code: \url{https://github.com/mir-group/flare} \cite{Vandermause2022}. Briefly, the geometry of local atomic environments (within a cutoff radius $r_{cut}$) is encoded in descriptors using the atomic cluster expansion (ACE) \cite{Drautz2019AtomicPotentials}. The normalized dot-product kernel is used to measure the similarity between atomic descriptors in a local energy prediction, which is then used in the construction of a sparse Gaussian process regression model (SGP). Additionally, the SGP provides an inherent mechanism to quantify predictive uncertainties for atomic forces, energies, and stresses, which can be used by the active learning algorithm to select \textit{ab initio} training data `on-the-fly' during an MD simulation. After each training is finished for the surrogate model, the SGP force field is mapped to a polynomial model \cite{Vandermause2022,Xie2021BayesianStanene} for high-efficiency in the MD simulation, being much faster than the SGP and without losing accuracy. 

For all parallel active learning trajectories, the ACE-B2 descriptor was employed. By using the second power of the normalized dot product kernel, `effective' 5-body interactions within each descriptor can be obtained, which is sufficiently complex for describing Pt with high accuracy \cite{owen2023complexity}. Maintaining consistency in notation with the original work of ACE \cite{Drautz2019AtomicPotentials}, ultimately, we chose $n_{\text{max}}=8$ for radial basis, $l_{\text{max}}=3$ for angular basis, and the cutoff radius $r_{cut} = 6 \rm{\AA}$ for all Pt-Pt, Pt-C, Pt-O, C-O, C-C, O-O and the inverse of such interactions. 

For more detailed descriptions of the FLARE active learning routine, we refer the reader to Ref. \cite{Vandermause2022}. The total number of DFT frames collected during active learning are provided in Suppl. Table 1, where 1,764 DFT calculations were performed across 6.84 nanoseconds of simulation time, and a maximum of 168 walltime hours for a given active learning trajectory. Including the final Allegro MLFF training, as is described in the next section, the MLFF took less than two weeks to construct. 

\subsection*{Equivariant MLFF training in Allegro}
The \textit{ab initio} frames collected during FLARE active learning were then combined and fed into the Allegro architecture \cite{Musaelian2023,10.1145/3581784.3627041}, where the final MLFF was trained. Following a fairly exhaustive grid search over model parameters, the final network employed two layers, 16 tensor features, multi-layer perceptron input dimensions of [64,128,256] for the scalar track, latent dimensions of [256,256,256], an angular resolution of $\mathcal{l}_{max}=4$, with an interaction cutoff of 8 \AA{} for all atom pairs. This cutoff was chosen due to single point energy calculations demonstrating the length-scale of such interactions in the system (e.g. CO above a full CO monolayer on Pt(111), where the energy and force labels only plateau at > 7 \AA{}), the evidence of which is provided in Supplementary Note 2. The model used a `default\_dtype` of 64 for floating point operations, and the final model was deployed with tensor float 32 turned off, which slightly reduces performance but maintains high accuracy especially in total energy calculations.

\subsection*{Molecular Dynamics Simulations in LAMMPS}
The final MD simulations were performed using a custom LAMMPS \cite{THOMPSON2022108171} pairstyle compiled for Allegro \cite{10.1145/3581784.3627041}. GPU acceleration was achieved with the Kokkos portability library \cite{CARTEREDWARDS20143202}. All simulations employed the Nos\'e-Hoover NVT ensemble with a time-step of $0.2$ fs. Velocities were randomly initialized for all simulations to a Boltzmann distribution centered at whatever desired temperature for the simulation (in units of Kelvin). Each NP was built using the minimized lattice constant of bulk Pt (4.03 \AA{}) as predicted by the MLFF in the Atomic Simulation Environment (ASE) \cite{Larsen_2017}. CO was introduced into the simulation box also using ASE.

\subsection*{Simulation Post-Processing in Ovito}
Trajectory analysis was performed using a custom Python code employing the OVITO API and all production MD trajectories were visually inspected with the OVITO graphical user interface \cite{ovito}. Python scripts were constructed around the OVITO pipeline in order to assess differences in Pt-Pt coordination (Fig. 3 \& Fig. 4) as a function of temperature and CO exposure, as well as any differences in the coordination environments that may be influenced by the nano-scale confinement of the nanoparticle shapes.

\subsection*{Chemicals}
Chloroplatinic acid hydrate (H$_2$PtCl$_6$·xH$_2$O, 99\%), 10–20 nm SiO$_2$ powder (Product No. 637238), poly(vinylpyrrolidone) (PVP, M$_w$ = 40000), ethylene glycol (EG, 99.8\%), and ethanol (99.5\%, anhydrous) were purchased from Sigma-Aldrich. NaOH (1 M) and acetone (99.5\%) were purchased from VWR. All reagent except the SiO$_2$ powder were used as received without further purification. The SiO$_2$ powder was used after thermal treatment at 973 K under air condition.

\subsection*{Pt/SiO$_2$ Sythesis}
A detailed synthesis procedure and characterization of the 2 nm Pt/SiO$_2$ used in this work are thoroughly described in our previous publications \cite{kim_interrogating_2024,kim2024well} and is briefly summarize below. The Pt colloidal synthesis was performed via the polyol method where 100 mg of H$_2$PtCl$_6$·xH2O, 2.5 mL of 1 M NaOH, and 20 mg of PVP were added to 10 mL of EG in a 50 mL three-neck round-bottom flask. The flask was evacuated at 273 K for 30 minutes while stirring vigorously (400 rpm), then heated to 353 K at a ramp rate of 10 K / min under vacuum and held at that temperature for 30 minutes. The temperature was then increased from 353 K to 473 K at a rate of 10 K / min, with Ar gas introduced at around 403 K. The solution was maintained at 473 K for 2 hours under Ar gas purging. After the reaction, the solution was cooled to room temperature. The colloidal suspension was diluted with 50 mL of acetone and centrifuged at 6500 rpm (2574.3 g) for 10 minutes twice. The Pt nanoparticles were then re-dispersed in 80 mL of ethanol and centrifuged at 2000 rpm (243.7 g) for 10 minutes. The supernatant contained the colloidal 2 nm Pt nanoparticles dispersed in ethanol.

The deposition of the Pt nanoparticles onto the SiO$_2$ support was performed by adding 3 mL of the Pt colloidal solution to 0.3 g of SiO$_2$ powder with vigorous stirring (400 rpm). The suspension was sonicated for 20 minutes and the solvent was evaporated under vacuum at 323 K overnight. The obtained Pt/SiO$_2$ catalyst powder was calcined at 773 K for 1 hour (ramp rate: 2 K / min) in air using a tube furnace to remove the majority of the carbonaceous capping layer.

\subsection*{Experimental Extended X-ray Absorption Fine Structure (EXAFS)}
XAS experiments at the Pt L3-edge were performed at the QAS beamline of the NSLSII. Approximately 35 mg of Pt/SiO$_2$ powder was pelletized using a 7 mm die and hand pelletizer. The sample was mounted to a Nashner-Adler cell \cite{Nashner} and measured in 0.6 \% CO in He at temperatures ranging from 298 K to 573 K. For each temperature condition, the EXAFS was collected in fluorescence mode using a PIPS detector 90 $^{\circ}$ from the incident beam direction. A 3 $\mu$m Zn filter was placed in front of the PIPS detector. Prior to exposure to CO, the sample was pretreated in 5 \% O2 in He at 573 K for 30 minutes and then in 5 \% H$_2$ for 30 minutes.

The EXAFS data was processed and fit quantitatively using the xraylarch \cite{Newville_2013} python package. The S$_0^2$ amplitude reduction factor was obtained from the fitting of the Pt foil, which was measured at the same experiment. The obtained value of 0.87 was used in all fittings. All data was fit well with one nearest-neighbor Pt-Pt photoelectron path. The k${^2}$-weighted data was Fourier transformed between 2-11.3 \AA{}$^{-1}$ with a Hanning window with dk=2. The R-space fitting range is 1.7 - 3.3 \AA{}. All EXAFS spectra were fit in parallel, where E0 was global and varied while individual coordination number, interatomic distance, Debye-Waller factor, and third cumulants were varied per spectrum.

\subsection*{Experimental Diffuse Reflectance Infrared Fourier Transform Spectroscopy (DRIFTS)}
DRIFTS experiments were performed in a low-volume reaction chamber (Harrick Scientific) equipped with ZnSe windows, mounted in a Bruker Invenio FT-IR spectrometer using a Praying Mantine diffuse reflectance accessory (Harrick Scientific). The DRIFTS reactor was sequentially loaded with a 304 stainless-steel mesh (150 $\times$ 150 msh), approximately 140 mg SiC, and approximately 8 mg of sieved Pt/SiO$_2$. The thermal gradient between the catalyst surface temperature and the measured thermocouple was calibrated by an optical pyrometer using an emissivity of 0.95 \cite{kim_interrogating_2024}. All DRIFTS experiments used a total volumetric flow rate of 25 mL min$^{-1}$. The Praying Mantis diffuse reflectance accessory and FT-IR spectrometer was purged with dry N$_2$ produced from compressed air by a purge gas generator (Parker). Each absorbance scan was obtained by averaging 200 background and sample scans at a resolution of 4 cm$^{-1}$ using a liquid nitrogen cooled HgCdTe (MCT) detector. The background measurements were acquired after the catalyst sample was annealed at 350 $^{\circ}$C for 30 mins in 5 \% H$_2$ in Ar and cooled to the reported spectra temperature. The sample measurements were acquired by maintaining the reported spectra temperature in 0.6 \% CO until a stable DRIFT spectra was achieved.

\subsection*{Data Availability}
The CO-Pt MLFF, all \textit{ab initio} training data, the training scripts, and experimental data will be provided on the Materials Cloud Archive upon publication.

\subsection*{Author Contributions} 
C.J.O.\ created the data set using FLARE active-learning, performed all MLFF training in Allegro, MD simulations in LAMMPS, postprocessing using the OVITO python API. N.M., C.R.O., and R.S. performed EXAFS measurements. C.R.O.\ performed DRIFTS measurements and analysis. N.M. performed EXAFS analysis and developed ML-MD trajectory analysis tools using the OVITO Python API.  C.J.O., N.M., and C.R.O.\ jointly created all figures, completed post-processing of the data, and co-led the writing of the manuscript, with contributions of all authors. T-S.K.\ synthesized the Pt/SiO$_2$ nanoparticle catalysts. C.Y.X.\ aided C.J.O.\ in validation of the MLFF, and helped write the length-scale interaction discussion provided in the SI. C.R.\ supervised all aspects of Pt/SiO$_2$ nanoparticle synthesis, and the DRIFTS data collection and analysis. A.I.F.\ supervised all aspects of the EXAFS data collection and analysis. B.K.\ supervised all aspects of the work. 

\subsection*{Acknowledgements} 
This work was primarily supported by the US Department of Energy, Office of Basic Energy Sciences Award No. DE-SC0022199 as well as by Robert Bosch LLC. C.J.O.\ was supported by the National Science Foundation Graduate Research Fellowship Program under Grant No. (DGE1745303). This research used resources of the National Energy Research Scientific Computing Center (NERSC), a DOE Office of Science User Facility supported by the Office of Science of the U.S. Department of Energy under Contract No. DE-AC02-05CH11231 using NERSC award BES-ERCAP0024206. An award for computer time was provided by the U.S. Department of Energy’s (DOE) Innovative and Novel Computational Impact on Theory and Experiment (INCITE) Program. This research used supporting resources at the Argonne and the Oak Ridge Leadership Computing Facilities. The Argonne Leadership Computing Facility at Argonne National Laboratory is supported by the Office of Science of the U.S. DOE under Contract No. DE-AC02-06CH11357. The Oak Ridge Leadership Computing Facility at the Oak Ridge National Laboratory is supported by the Office of Science of the U.S. DOE under Contract No. DE-AC05-00OR22725. Additional computational resources were provided by the FAS Division of Science Research Computing Group at Harvard University and the Theory and Computation facility of the Center for Functional Nanomaterials (CFN), which is a U.S. Department of Energy Office of Science User Facility, at Brookhaven National Laboratory under Contract No. DE-SC0012704. C.R. gratefully acknowledges the Rowland Fellowship through the Rowland Institute at Harvard.

\subsection*{Competing interests}
The authors declare no competing interests.

\subsection*{References}
\bibliography{bib.bib}

\begin{thebibliography}{10}

\bibitem{Snitkoff-Sol_Friedman_Honig_Yurko_Kozhushner_Zachman_Zelenay_Bond_Elbaz_2022}
R.~Z. Snitkoff-Sol \emph{et~al.}, Quantifying the electrochemical active site
  density of precious metal-free catalysts in situ in fuel cells, {\em Nature
  Catalysis}, 5(2), 163–170, (2022).

\bibitem{Vogt_Weckhuysen_2022}
C.~Vogt and B.~M. Weckhuysen, The concept of active site in heterogeneous
  catalysis, {\em Nature Reviews Chemistry}, 6(2), 89–111, (2022).

\bibitem{Somorjai_1985}
G.~A. Somorjai, Surface science and catalysis, {\em Science}, 227(4689),
  902–908, (1985).

\bibitem{frenkel_view_2001}
A.~I. Frenkel, C.~W. Hills, and R.~G. Nuzzo, A {View} from the {Inside}:
  {Complexity} in the {Atomic} {Scale} {Ordering} of {Supported} {Metal}
  {Nanoparticles}, {\em The Journal of Physical Chemistry B}, 105(51),
  12689--12703, (2001).

\bibitem{mostafa2010shape}
S.~Mostafa \emph{et~al.}, Shape-dependent catalytic properties of pt
  nanoparticles, {\em Journal of the American Chemical Society}, 132(44),
  15714--15719, (2010).

\bibitem{Nørskov_Bligaard_Rossmeisl_Christensen_2009}
J.~K. Nørskov, T.~Bligaard, J.~Rossmeisl, and C.~H. Christensen, Towards the
  computational design of solid catalysts, {\em Nature Chemistry}, 1(1),
  37–46, (2009).

\bibitem{Beck_Paunovic_VanBokhoven_2023}
A.~Beck, V.~Paunović, and J.~A. Van~Bokhoven, Identifying and avoiding dead
  ends in the characterization of heterogeneous catalysts at the gas–solid
  interface, {\em Nature Catalysis}, 6(10), 873–884, (2023).

\bibitem{Bruix_Margraf_Andersen_Reuter_2019}
A.~Bruix, J.~T. Margraf, M.~Andersen, and K.~Reuter, First-principles-based
  multiscale modelling of heterogeneous catalysis, {\em Nature Catalysis},
  2(8), 659–670, (2019).

\bibitem{PhysRevB.78.121404}
F.~Vila, J.~J. Rehr, J.~Kas, R.~G. Nuzzo, and A.~I. Frenkel, Dynamic structure
  in supported pt nanoclusters: Real-time density functional theory and x-ray
  spectroscopy simulations, {\em Phys. Rev. B}, 78, 121404, (2008).

\bibitem{vila2017anomalous}
F.~D. Vila, J.~J. Rehr, R.~G. Nuzzo, and A.~I. Frenkel, Anomalous structural
  disorder in supported pt nanoparticles, {\em The journal of physical
  chemistry letters}, 8(14), 3284--3288, (2017).

\bibitem{Liu_Shih_Deng_Ojha_Chen_Luo_McCrum_Koper_Greeley_Zeng_2024}
G.~Liu \emph{et~al.}, Site-specific reactivity of stepped pt surfaces driven by
  stress release, {\em Nature}, 626(8001), 1005–1010, (2024).

\bibitem{Reece_Redekop_Karakalos_Friend_Madix_2018}
C.~Reece, E.~A. Redekop, S.~Karakalos, C.~M. Friend, and R.~J. Madix, Crossing
  the great divide between single-crystal reactivity and actual catalyst
  selectivity with pressure transients, {\em Nature Catalysis}, 1(11),
  852–859, (2018).

\bibitem{Seh_Kibsgaard_Dickens_Chorkendorff_Nørskov_Jaramillo_2017}
Z.~W. Seh \emph{et~al.}, Combining theory and experiment in electrocatalysis:
  Insights into materials design, {\em Science}, 355(6321), eaad4998, (2017).

\bibitem{Marcella2022}
N.~Marcella \emph{et~al.}, Decoding reactive structures in dilute alloy
  catalysts, {\em Nature Communications}, 13(1), 832, (2022).

\bibitem{Mou_Pillai_Wang_Wan_Han_Schweitzer_Che_Xin_2023}
T.~Mou \emph{et~al.}, Bridging the complexity gap in computational
  heterogeneous catalysis with machine learning, {\em Nature Catalysis}, 6(2),
  122–136, (2023).

\bibitem{Somorjai_1992}
G.~A. Somorjai, The experimental evidence of the role of surface restructuring
  during catalytic reactions, {\em Catalysis Letters}, 12(1), 17–34, (1992).

\bibitem{Somorjai_Carrazza_1986}
G.~A. Somorjai and J.~Carrazza, Structure sensitivity of catalytic reactions,
  {\em Industrial \& Engineering Chemistry Fundamentals}, 25(1), 63–69,
  (1986).

\bibitem{huang2008coordination}
W.~Huang \emph{et~al.}, Coordination-dependent surface atomic contraction in
  nanocrystals revealed by coherent diffraction, {\em Nature materials}, 7(4),
  308--313, (2008).

\bibitem{Xu_Papanikolaou_Lechner_Je_Somorjai_Salmeron_Mavrikakis_2023}
L.~Xu \emph{et~al.}, Formation of active sites on transition metals through
  reaction-driven migration of surface atoms, {\em Science}, 380(6640),
  70–76, (2023).

\bibitem{Eren_Zherebetskyy_Patera_Wu_Bluhm_Africh_Wang_Somorjai_Salmeron_2016}
B.~Eren \emph{et~al.}, Activation of cu(111) surface by decomposition into
  nanoclusters driven by co adsorption, {\em Science}, 351(6272), 475–478,
  (2016).

\bibitem{Eren_Torres_Karslıoğlu_Liu_Wu_Stacchiola_Bluhm_Somorjai_Salmeron_2018}
B.~Eren \emph{et~al.}, Structure of copper–cobalt surface alloys in
  equilibrium with carbon monoxide gas, {\em Journal of the American Chemical
  Society}, 140(21), 6575–6581, (2018).

\bibitem{Roiaz_Falivene_Rameshan_Cavallo_Kozlov_Rupprechter_2019}
M.~Roiaz \emph{et~al.}, Roughening of copper (100) at elevated co pressure: Cu
  adatom and cluster formation enable co dissociation, {\em The Journal of
  Physical Chemistry C}, 123(13), 8112–8121, (2019).

\bibitem{Tao_Dag_Wang_Liu_Butcher_Bluhm_Salmeron_Somorjai_2010}
F.~Tao \emph{et~al.}, Break-up of stepped platinum catalyst surfaces by high co
  coverage, {\em Science}, 327(5967), 850–853, (2010).

\bibitem{li_noncrystalline--crystalline_2013}
L.~Li \emph{et~al.}, Noncrystalline-to-{Crystalline} {Transformations} in {Pt}
  {Nanoparticles}, {\em Journal of the American Chemical Society}, 135(35),
  13062--13072, (2013).

\bibitem{Vila2008}
F.~Vila, J.~J. Rehr, J.~Kas, R.~G. Nuzzo, and A.~I. Frenkel, Dynamic structure
  in supported pt nanoclusters: Real-time density functional theory and x-ray
  spectroscopy simulations, {\em Phys. Rev. B}, 78, 121404, (2008).

\bibitem{Erickson2014}
E.~M. Erickson \emph{et~al.}, A comparison of atomistic and continuum
  approaches to the study of bonding dynamics in electrocatalysis:
  Microcantilever stress and in situ exafs observations of platinum bond
  expansion due to oxygen adsorption during the oxygen reduction reaction, {\em
  Analytical Chemistry}, 86(16), 8368--8375, (2014).
\newblock PMID: 25066179.

\bibitem{Li2021}
Y.~Li \emph{et~al.}, Dynamic structure of active sites in ceria-supported pt
  catalysts for the water gas shift reaction, {\em Nature Communications},
  12(1), 914, (2021).

\bibitem{Vandermause2020}
J.~Vandermause \emph{et~al.}, On-the-fly active learning of interpretable
  bayesian force fields for atomistic rare events, {\em npj Computational
  Materials}, 6(1), 20, (2020).

\bibitem{Vandermause2022}
J.~Vandermause, Y.~Xie, J.~S. Lim, C.~J. Owen, and B.~Kozinsky, Active learning
  of reactive bayesian force fields applied to heterogeneous catalysis dynamics
  of h/pt, {\em Nature Communications}, 13(1), 5183, (2022).

\bibitem{Batzner2021E3-EquivariantPotentials}
S.~Batzner \emph{et~al.}, {E(3)-Equivariant graph neural networks for
  data-efficient and accurate interatomic potentials}, {\em Nature
  Communications 2022 13:1}, 13(1), 1--11, (2021).

\bibitem{Musaelian2023}
A.~Musaelian \emph{et~al.}, Learning local equivariant representations for
  large-scale atomistic dynamics, {\em Nature Communications}, 14(1), 579,
  (2023).

\bibitem{NEURIPS2022_4a36c3c5}
I.~Batatia, D.~P. Kovacs, G.~Simm, C.~Ortner, and G.~Csanyi.
\newblock Mace: Higher order equivariant message passing neural networks for
  fast and accurate force fields.
\newblock In S.~Koyejo \emph{et~al.}, editors, {\em Advances in Neural
  Information Processing Systems}, volume~35, pp. 11423--11436. Curran
  Associates, Inc., (2022).

\bibitem{Shapeev2016MomentPotentials}
A.~V. Shapeev, Moment tensor potentials: a class of systematically improvable
  interatomic potentials, {\em Multiscale Modeling \& Simulation}, 14(3),
  1153--1173, (2016).

\bibitem{owen2023unraveling}
C.~J. Owen \emph{et~al.}, {\em arXiv:2306.00901}, (2023).

\bibitem{Hollins_1992}
P.~Hollins, The influence of surface defects on the infrared spectra of
  adsorbed species, {\em Surface Science Reports}, 16(2), 51–94, (1992).

\bibitem{Kale_Christopher_2016}
M.~J. Kale and P.~Christopher, Utilizing quantitative in situ ftir spectroscopy
  to identify well-coordinated pt atoms as the active site for co oxidation on
  al2o3-supported pt catalysts, {\em ACS Catalysis}, 6(8), 5599–5609, (2016).

\bibitem{kim_interrogating_2024}
T.-S. Kim, C.~R. O’Connor, and C.~Reece, Interrogating site dependent
  kinetics over {SiO2}-supported {Pt} nanoparticles, {\em Nature
  Communications}, 15(1), 2074, (2024).

\bibitem{crossley1980adsorbate}
A.~Crossley and D.~A. King, Adsorbate island dimensions and interaction
  energies from vibrational spectra: Co on pt $\{$001$\}$ and pt $\{$111$\}$,
  {\em Surface Science}, 95(1), 131--155, (1980).

\bibitem{Frenkel_Hills_Nuzzo_2001}
A.~I. Frenkel, C.~W. Hills, and R.~G. Nuzzo, A view from the inside: Complexity
  in the atomic scale ordering of supported metal nanoparticles, {\em The
  Journal of Physical Chemistry B}, 105(51), 12689–12703, (2001).

\bibitem{FrenkelRehr1993}
A.~I. Frenkel and J.~J. Rehr, Thermal expansion and x-ray-absorption
  fine-structure cumulants, {\em Phys. Rev. B}, 48, 585--588, (1993).

\bibitem{Owen2024_au}
C.~J. Owen, Y.~Xie, A.~Johansson, L.~Sun, and B.~Kozinsky, Low-index mesoscopic
  surface reconstructions of au surfaces using bayesian force fields, {\em
  Nature Communications}, 15(1), 3790, (2024).

\bibitem{owen2024unbiased}
C.~J. Owen \emph{et~al.}, Unbiased atomistic predictions of crystal dislocation
  dynamics using bayesian force fields, {\em arXiv preprint arXiv:2401.04359},
  (2024).

\bibitem{Johansson2022Micron-scaleLearning}
A.~Johansson \emph{et~al.}, {Micron-scale heterogeneous catalysis with Bayesian
  force fields from first principles and active learning}, {\em arXiv preprint
  arXiv:2204.12573}, (2022).

\bibitem{Xie2023Uncertainty-awareSiC}
Y.~Xie \emph{et~al.}, Uncertainty-aware molecular dynamics from bayesian active
  learning for phase transformations and thermal transport in sic, {\em npj
  Computational Materials}, 9(1), 36, (2023).

\bibitem{Xie2021BayesianStanene}
Y.~Xie, J.~Vandermause, L.~Sun, A.~Cepellotti, and B.~Kozinsky, {Bayesian force
  fields from active learning for simulation of inter-dimensional
  transformation of stanene}, {\em npj Computational Materials}, 7(1), 1--10,
  (2021).

\bibitem{10.1145/3581784.3627041}
A.~Musaelian, A.~Johansson, S.~Batzner, and B.~Kozinsky.
\newblock Scaling the leading accuracy of deep equivariant models to
  biomolecular simulations of realistic size.
\newblock In {\em Proceedings of the International Conference for High
  Performance Computing, Networking, Storage and Analysis}, SC '23, New York,
  NY, USA, (2023). Association for Computing Machinery.

\bibitem{doi:10.1021/jacs.2c10179}
V.~Sumaria, L.~Nguyen, F.~F. Tao, and P.~Sautet, Atomic-scale mechanism of
  platinum catalyst restructuring under a pressure of reactant gas, {\em
  Journal of the American Chemical Society}, 145(1), 392--401, (2023).
\newblock PMID: 36548635.

\bibitem{D1SC03827C}
V.~Sumaria and P.~Sautet, Co organization at ambient pressure on stepped pt
  surfaces: first principles modeling accelerated by neural networks, {\em
  Chem. Sci.}, 12, 15543--15555, (2021).

\bibitem{doi:10.1021/acscatal.0c01971}
V.~Sumaria, L.~Nguyen, F.~F. Tao, and P.~Sautet, Optimal packing of co at a
  high coverage on pt(100) and pt(111) surfaces, {\em ACS Catalysis}, 10(16),
  9533--9544, (2020).

\bibitem{Ahmadi_Timoshenko_Behafarid_Roldan_Cuenya_2019}
M.~Ahmadi, J.~Timoshenko, F.~Behafarid, and B.~Roldan~Cuenya, Tuning the
  structure of pt nanoparticles through support interactions: An in situ
  polarized x-ray absorption study coupled with atomistic simulations, {\em The
  Journal of Physical Chemistry C}, 123(16), 10666–10676, (2019).

\bibitem{Simonsen_Chorkendorff_Dahl_Skoglundh_Sehested_Helveg_2011}
S.~B. Simonsen \emph{et~al.}, Ostwald ripening in a pt/sio2 model catalyst
  studied by in situ tem, {\em Journal of Catalysis}, 281(1), 147–155,
  (2011).

\bibitem{Wang_fragmentation_2024}
H.~Wang \emph{et~al.}, Unravelling the origin of reaction-driven aggregation
  and fragmentation of atomically dispersed pt catalyst on ceria support, {\em
  Nanoscale}, pp.~--, (2024).

\bibitem{kalhara2018co}
G.~K. Kalhara~Gunasooriya and M.~Saeys, Co adsorption site preference on
  platinum: charge is the essence, {\em ACS Catalysis}, 8(5), 3770--3774,
  (2018).

\bibitem{catal8100450}
C.~Yu \emph{et~al.}, H2 thermal desorption spectra on pt(111): A density
  functional theory and kinetic monte carlo simulation study, {\em Catalysts},
  8(10), (2018).

\bibitem{ovito}
A.~Stukowski, {Visualization and analysis of atomistic simulation data with
  OVITO-the Open Visualization Tool}, {\em Modelling and Simulation in
  Materials Science and Engineering}, {18}({1}), ({2010}).

\bibitem{Yang2023}
M.~Yang, U.~Raucci, and M.~Parrinello, Reactant-induced dynamics of lithium
  imide surfaces during the ammonia decomposition process, {\em Nature
  Catalysis}, 6(9), 829--836, (2023).

\bibitem{doi:10.1021/acs.jpca.6b06770}
Y.~K. Shin, L.~Gai, S.~Raman, and A.~C.~T. van Duin, Development of a reaxff
  reactive force field for the pt–ni alloy catalyst, {\em The Journal of
  Physical Chemistry A}, 120(41), 8044--8055, (2016).
\newblock PMID: 27670674.

\bibitem{PhysRevB.69.161401}
S.~E. Mason, I.~Grinberg, and A.~M. Rappe, First-principles extrapolation
  method for accurate co adsorption energies on metal surfaces, {\em Phys. Rev.
  B}, 69, 161401, (2004).

\bibitem{doi:10.1021/jp002302t}
P.~J. Feibelman \emph{et~al.}, The co/pt(111) puzzle, {\em The Journal of
  Physical Chemistry B}, 105(18), 4018--4025, (2001).

\bibitem{Hammer1995WhyMetals}
B.~Hammer and J.~K. Norskov, {Why gold is the noblest of all the metals}, {\em
  Nature}, 376(6537), 238--240, (1995).

\bibitem{Kresse1996}
G.~Kresse and J.~Furthm{\"{u}}ller, {Efficiency of ab-initio total energy
  calculations for metals and semiconductors using a plane-wave basis set},
  {\em Computational Materials Science}, 6(1), 15--50, (1996).

\bibitem{THOMPSON2022108171}
A.~P. Thompson \emph{et~al.}, Lammps - a flexible simulation tool for
  particle-based materials modeling at the atomic, meso, and continuum scales,
  {\em Computer Physics Communications}, 271, 108171, (2022).

\bibitem{Drautz2019AtomicPotentials}
R.~Drautz, {Atomic cluster expansion for accurate and transferable interatomic
  potentials}, {\em Physical Review B}, 99(1), 014104, (2019).

\bibitem{owen2023complexity}
C.~J. Owen \emph{et~al.}, Complexity of many-body interactions in transition
  metals via machine-learned force fields from the tm23 data set, {\em arXiv
  preprint arXiv:2302.12993}, (2023).

\bibitem{CARTEREDWARDS20143202}
H.~{Carter Edwards}, C.~R. Trott, and D.~Sunderland, Kokkos: Enabling manycore
  performance portability through polymorphic memory access patterns, {\em
  Journal of Parallel and Distributed Computing}, 74(12), 3202--3216, (2014).
\newblock Domain-Specific Languages and High-Level Frameworks for
  High-Performance Computing.

\bibitem{Larsen_2017}
A.~H. Larsen \emph{et~al.}, The atomic simulation environment—a python
  library for working with atoms, {\em Journal of Physics: Condensed Matter},
  29(27), 273002, (2017).

\bibitem{kim2024well}
T.-S. Kim, C.~R. O’Connor, S.~Le, and C.~Reece, A well-defined supported pt
  nanoparticle catalyst for heterogeneous catalytic surface science, {\em
  Journal of Materials Chemistry A}, (2024).

\bibitem{Nashner}
M.~S. Nashner, A.~I. Frenkel, D.~L. Adler, J.~R. Shapley, and R.~G. Nuzzo,
  Structural characterization of carbon-supported platinum−ruthenium
  nanoparticles from the molecular cluster precursor ptru5c(co)16, {\em Journal
  of the American Chemical Society}, 119(33), 7760--7771, (1997).

\bibitem{Newville_2013}
M.~Newville, Larch: An analysis package for xafs and related spectroscopies,
  {\em Journal of Physics: Conference Series}, 430(1), 012007, (2013).

\end{thebibliography}


\begin{thebibliography}{}

\end{thebibliography}



ENTRY
  { address
    author
    booktitle
    chapter
    edition
    editor
    howpublished
    institution
    journal
    key
    month
    note
    number
    organization
    pages
    publisher
    school
    series
    title
    type
    volume
    year
  }
  {}
  { label }

INTEGERS { output.state before.all mid.sentence after.sentence after.block }

FUNCTION {init.state.consts}
{ #0 'before.all :=
  #1 'mid.sentence :=
  #2 'after.sentence :=
  #3 'after.block :=
}

STRINGS { s t }

FUNCTION {output.nonnull}
{ 's :=
  output.state mid.sentence =
    { ", " * write$ }
    { output.state after.block =
        { add.period$ write$
          newline$
          "\newblock " write$
        }
        { output.state before.all =
            'write$
            { add.period$ " " * write$ }
          if$
        }
      if$
      mid.sentence 'output.state :=
    }
  if$
  s
}

FUNCTION {output}
{ duplicate$ empty$
    'pop$
    'output.nonnull
  if$
}

FUNCTION {output.check}
{ 't :=
  duplicate$ empty$
    { pop$ "empty " t * " in " * cite$ * warning$ }
    'output.nonnull
  if$
}

FUNCTION {output.bibitem}
{ newline$
  "\bibitem{" write$
  cite$ write$
  "}" write$
  newline$
  ""
  before.all 'output.state :=
}

FUNCTION {fin.entry}
{ add.period$
  write$
  newline$
}

FUNCTION {new.block}
{ output.state before.all =
    'skip$
    { after.block 'output.state := }
  if$
}

FUNCTION {new.sentence}
{ output.state after.block =
    'skip$
    { output.state before.all =
        'skip$
        { after.sentence 'output.state := }
      if$
    }
  if$
}

FUNCTION {not}
{   { #0 }
    { #1 }
  if$
}

FUNCTION {and}
{   'skip$
    { pop$ #0 }
  if$
}

FUNCTION {or}
{   { pop$ #1 }
    'skip$
  if$
}

FUNCTION {new.block.checka}
{ empty$
    'skip$
    'new.block
  if$
}

FUNCTION {new.block.checkb}
{ empty$
  swap$ empty$
  and
    'skip$
    'new.block
  if$
}

FUNCTION {new.sentence.checka}
{ empty$
    'skip$
    'new.sentence
  if$
}

FUNCTION {new.sentence.checkb}
{ empty$
  swap$ empty$
  and
    'skip$
    'new.sentence
  if$
}

FUNCTION {field.or.null}
{ duplicate$ empty$
    { pop$ "" }
    'skip$
  if$
}

FUNCTION {emphasize}
{ duplicate$ empty$
    { pop$ "" }
    { "{\em " swap$ * "}" * }
  if$
}

INTEGERS { nameptr namesleft numnames }


FUNCTION {format.names}
{ 's :=
  #1 'nameptr :=
  s num.names$ 'numnames :=
  numnames 'namesleft :=
  numnames #5 >
    { s #1 "{f.~}{vv~}{ll}{, jj}" format.name$
      " \emph{et~al.}" * }
    {
      { namesleft #0 > }
      { s nameptr "{f.~}{vv~}{ll}{, jj}" format.name$ 't :=
        nameptr #1 >
          { namesleft #1 >
              { ", " * t * }
              { numnames #2 >
                  { "," * }
                  'skip$
                if$
                t "others" =
                  { " \emph{et~al}." * }
                  { " and " * t * }
                if$
              }
            if$
          }
          't
        if$
        nameptr #1 + 'nameptr :=
        namesleft #1 - 'namesleft :=
      }
    while$
  }
  if$
}


FUNCTION {format.authors}
{ author empty$
    { "" }
    { author format.names }
  if$
}

FUNCTION {format.editors}
{ editor empty$
    { "" }
    { editor format.names
      editor num.names$ #1 >
        { ", editors" * }
        { ", editor" * }
      if$
    }
  if$
}

FUNCTION {format.title}
{ title empty$
    { "" }
    { title "t" change.case$ }
  if$
}

FUNCTION {n.dashify}
{ 't :=
  ""
    { t empty$ not }
    { t #1 #1 substring$ "-" =
        { t #1 #2 substring$ "--" = not
            { "--" *
              t #2 global.max$ substring$ 't :=
            }
            {   { t #1 #1 substring$ "-" = }
                { "-" *
                  t #2 global.max$ substring$ 't :=
                }
              while$
            }
          if$
        }
        { t #1 #1 substring$ *
          t #2 global.max$ substring$ 't :=
        }
      if$
    }
  while$
}

FUNCTION {format.date}
{ "("  year ")" * *
}

FUNCTION {format.btitle}
{ title emphasize
}

FUNCTION {tie.or.space.connect}
{ duplicate$ text.length$ #3 <
    { "~" }
    { " " }
  if$
  swap$ * *
}

FUNCTION {either.or.check}
{ empty$
    'pop$
    { "can't use both " swap$ * " fields in " * cite$ * warning$ }
  if$
}

FUNCTION {format.bvolume}
{ volume empty$
    { "" }
    { "volume" volume tie.or.space.connect
      series empty$
        'skip$
        { " of " * series emphasize * }
      if$
      "volume and number" number either.or.check
    }
  if$
}

FUNCTION {format.number.series}
{ volume empty$
    { number empty$
        { series field.or.null }
        { output.state mid.sentence =
            { "number" }
            { "Number" }
          if$
          number tie.or.space.connect
          series empty$
            { "there's a number but no series in " cite$ * warning$ }
            { " in " * series * }
          if$
        }
      if$
    }
    { "" }
  if$
}

FUNCTION {format.edition}
{ edition empty$
    { "" }
    { output.state mid.sentence =
        { edition "l" change.case$ " edition" * }
        { edition "t" change.case$ " edition" * }
      if$
    }
  if$
}

INTEGERS { multiresult }

FUNCTION {multi.page.check}
{ 't :=
  #0 'multiresult :=
    { multiresult not
      t empty$ not
      and
    }
    { t #1 #1 substring$
      duplicate$ "-" =
      swap$ duplicate$ "," =
      swap$ "+" =
      or or
        { #1 'multiresult := }
        { t #2 global.max$ substring$ 't := }
      if$
    }
  while$
  multiresult
}

FUNCTION {format.pages}
{ pages empty$
    { "" }
    { pages multi.page.check
        { "pp." pages n.dashify tie.or.space.connect }
        { "pp." pages tie.or.space.connect }
      if$
    }
  if$
}

FUNCTION {format.vol.num.pages}
{ volume field.or.null
  number empty$
    'skip$
    { "(" number * ")" * *
      volume empty$
        { "there's a number but no volume in " cite$ * warning$ }
        'skip$
      if$
    }
  if$
  pages empty$
    'skip$
    { duplicate$ empty$
        { pop$ format.pages }
        { ", " * pages n.dashify * }
      if$
    }
  if$
}

FUNCTION {format.chapter.pages}
{ chapter empty$
    'format.pages
    { type empty$
        { "chapter" }
        { type "l" change.case$ }
      if$
      chapter tie.or.space.connect
      pages empty$
        'skip$
        { ", " * format.pages * }
      if$
    }
  if$
}

FUNCTION {format.in.ed.booktitle}
{ booktitle empty$
    { "" }
    { editor empty$
        { "In " booktitle emphasize * }
        { "In " format.editors * ", " * booktitle emphasize * }
      if$
    }
  if$
}

FUNCTION {empty.misc.check}
{ author empty$ title empty$ howpublished empty$
  month empty$ year empty$ note empty$
  and and and and and
    { "all relevant fields are empty in " cite$ * warning$ }
    'skip$
  if$
}

FUNCTION {format.thesis.type}
{ type empty$
    'skip$
    { pop$
      type "t" change.case$
    }
  if$
}

FUNCTION {format.tr.number}
{ type empty$
    { "Technical Report" }
    'type
  if$
  number empty$
    { "t" change.case$ }
    { number tie.or.space.connect }
  if$
}

FUNCTION {format.article.crossref}
{ key empty$
    { journal empty$
        { "need key or journal for " cite$ * " to crossref " * crossref *
          warning$
          ""
        }
        { "In {\em " journal * "\/}" * }
      if$
    }
    { "In " key * }
  if$
  " \cite{" * crossref * "}" *
}

FUNCTION {format.crossref.editor}
{ editor #1 "{vv~}{ll}" format.name$
  editor num.names$ duplicate$
  #2 >
    { pop$ " et~al." * }
    { #2 <
        'skip$
        { editor #2 "{ff }{vv }{ll}{ jj}" format.name$ "others" =
            { " et~al." * }
            { " and " * editor #2 "{vv~}{ll}" format.name$ * }
          if$
        }
      if$
    }
  if$
}

FUNCTION {format.book.crossref}
{ volume empty$
    { "empty volume in " cite$ * "'s crossref of " * crossref * warning$
      "In "
    }
    { "Volume" volume tie.or.space.connect
      " of " *
    }
  if$
  editor empty$
  editor field.or.null author field.or.null =
  or
    { key empty$
        { series empty$
            { "need editor, key, or series for " cite$ * " to crossref " *
              crossref * warning$
              "" *
            }
            { "{\em " * series * "\/}" * }
          if$
        }
        { key * }
      if$
    }
    { format.crossref.editor * }
  if$
  " \cite{" * crossref * "}" *
}

FUNCTION {format.incoll.inproc.crossref}
{ editor empty$
  editor field.or.null author field.or.null =
  or
    { key empty$
        { booktitle empty$
            { "need editor, key, or booktitle for " cite$ * " to crossref " *
              crossref * warning$
              ""
            }
            { "In {\em " booktitle * "\/}" * }
          if$
        }
        { "In " key * }
      if$
    }
    { "In " format.crossref.editor * }
  if$
  " \cite{" * crossref * "}" *
}

FUNCTION {article}
{ output.bibitem
  format.authors "author" output.check
  format.title "title" output.check
  crossref missing$
    { journal emphasize "journal" output.check
      format.vol.num.pages output
      format.date "year" output.check
    }
    { format.article.crossref output.nonnull
      format.pages output
    }
  if$
  new.block
  note output
  fin.entry
}

FUNCTION {book}
{ output.bibitem
  author empty$
    { format.editors "author and editor" output.check }
    { format.authors output.nonnull
      crossref missing$
        { "author and editor" editor either.or.check }
        'skip$
      if$
    }
  if$
  new.block
  format.btitle "title" output.check
  crossref missing$
    { format.bvolume output
      new.block
      format.number.series output
      new.sentence
      publisher "publisher" output.check
      address output
    }
    { new.block
      format.book.crossref output.nonnull
    }
  if$
  format.edition output
  format.date "year" output.check
  new.block
  note output
  fin.entry
}

FUNCTION {booklet}
{ output.bibitem
  format.authors output
  new.block
  format.title "title" output.check
  howpublished address new.block.checkb
  howpublished output
  address output
  format.date output
  new.block
  note output
  fin.entry
}

FUNCTION {inbook}
{ output.bibitem
  author empty$
    { format.editors "author and editor" output.check }
    { format.authors output.nonnull
      crossref missing$
        { "author and editor" editor either.or.check }
        'skip$
      if$
    }
  if$
  new.block
  format.btitle "title" output.check
  crossref missing$
    { format.bvolume output
      format.chapter.pages "chapter and pages" output.check
      new.block
      format.number.series output
      new.sentence
      publisher "publisher" output.check
      address output
    }
    { format.chapter.pages "chapter and pages" output.check
      new.block
      format.book.crossref output.nonnull
    }
  if$
  format.edition output
  format.date "year" output.check
  new.block
  note output
  fin.entry
}

FUNCTION {incollection}
{ output.bibitem
  format.authors "author" output.check
  new.block
  format.title "title" output.check
  new.block
  crossref missing$
    { format.in.ed.booktitle "booktitle" output.check
      format.bvolume output
      format.number.series output
      format.chapter.pages output
      new.sentence
      publisher "publisher" output.check
      address output
      format.edition output
      format.date "year" output.check
    }
    { format.incoll.inproc.crossref output.nonnull
      format.chapter.pages output
    }
  if$
  new.block
  note output
  fin.entry
}

FUNCTION {inproceedings}
{ output.bibitem
  format.authors "author" output.check
  new.block
  format.title "title" output.check
  new.block
  crossref missing$
    { format.in.ed.booktitle "booktitle" output.check
      format.bvolume output
      format.number.series output
      format.pages output
      address empty$
        { organization publisher new.sentence.checkb
          organization output
          publisher output
          format.date "year" output.check
        }
        { address output.nonnull
          format.date "year" output.check
          new.sentence
          organization output
          publisher output
        }
      if$
    }
    { format.incoll.inproc.crossref output.nonnull
      format.pages output
    }
  if$
  new.block
  note output
  fin.entry
}

FUNCTION {conference} { inproceedings }

FUNCTION {manual}
{ output.bibitem
  author empty$
    { organization empty$
        'skip$
        { organization output.nonnull
          address output
        }
      if$
    }
    { format.authors output.nonnull }
  if$
  new.block
  format.btitle "title" output.check
  author empty$
    { organization empty$
        { address new.block.checka
          address output
        }
        'skip$
      if$
    }
    { organization address new.block.checkb
      organization output
      address output
    }
  if$
  format.edition output
  format.date output
  new.block
  note output
  fin.entry
}

FUNCTION {mastersthesis}
{ output.bibitem
  format.authors "author" output.check
  new.block
  format.title "title" output.check
  new.block
  "Master's thesis" format.thesis.type output.nonnull
  school "school" output.check
  address output
  format.date "year" output.check
  new.block
  note output
  fin.entry
}

FUNCTION {misc}
{ output.bibitem
  format.authors output
  title howpublished new.block.checkb
  format.title output
  howpublished new.block.checka
  howpublished output
  format.date output
  new.block
  note output
  fin.entry
  empty.misc.check
}

FUNCTION {phdthesis}
{ output.bibitem
  format.authors "author" output.check
  new.block
  format.btitle "title" output.check
  new.block
  "PhD thesis" format.thesis.type output.nonnull
  school "school" output.check
  address output
  format.date "year" output.check
  new.block
  note output
  fin.entry
}

FUNCTION {proceedings}
{ output.bibitem
  editor empty$
    { organization output }
    { format.editors output.nonnull }
  if$
  new.block
  format.btitle "title" output.check
  format.bvolume output
  format.number.series output
  address empty$
    { editor empty$
        { publisher new.sentence.checka }
        { organization publisher new.sentence.checkb
          organization output
        }
      if$
      publisher output
      format.date "year" output.check
    }
    { address output.nonnull
      format.date "year" output.check
      new.sentence
      editor empty$
        'skip$
        { organization output }
      if$
      publisher output
    }
  if$
  new.block
  note output
  fin.entry
}

FUNCTION {techreport}
{ output.bibitem
  format.authors "author" output.check
  new.block
  format.title "title" output.check
  new.block
  format.tr.number output.nonnull
  institution "institution" output.check
  address output
  format.date "year" output.check
  new.block
  note output
  fin.entry
}

FUNCTION {unpublished}
{ output.bibitem
  format.authors "author" output.check
  new.block
  format.title "title" output.check
  new.block
  note "note" output.check
  format.date output
  fin.entry
}

FUNCTION {default.type} { misc }

MACRO {jan} {"January"}

MACRO {feb} {"February"}

MACRO {mar} {"March"}

MACRO {apr} {"April"}

MACRO {may} {"May"}

MACRO {jun} {"June"}

MACRO {jul} {"July"}

MACRO {aug} {"August"}

MACRO {sep} {"September"}

MACRO {oct} {"October"}

MACRO {nov} {"November"}

MACRO {dec} {"December"}

MACRO {acmcs} {"ACM Computing Surveys"}

MACRO {acta} {"Acta Informatica"}

MACRO {cacm} {"Communications of the ACM"}

MACRO {ibmjrd} {"IBM Journal of Research and Development"}

MACRO {ibmsj} {"IBM Systems Journal"}

MACRO {ieeese} {"IEEE Transactions on Software Engineering"}

MACRO {ieeetc} {"IEEE Transactions on Computers"}

MACRO {ieeetcad}
 {"IEEE Transactions on Computer-Aided Design of Integrated Circuits"}

MACRO {ipl} {"Information Processing Letters"}

MACRO {jacm} {"Journal of the ACM"}

MACRO {jcss} {"Journal of Computer and System Sciences"}

MACRO {scp} {"Science of Computer Programming"}

MACRO {sicomp} {"SIAM Journal on Computing"}

MACRO {tocs} {"ACM Transactions on Computer Systems"}

MACRO {tods} {"ACM Transactions on Database Systems"}

MACRO {tog} {"ACM Transactions on Graphics"}

MACRO {toms} {"ACM Transactions on Mathematical Software"}

MACRO {toois} {"ACM Transactions on Office Information Systems"}

MACRO {toplas} {"ACM Transactions on Programming Languages and Systems"}

MACRO {tcs} {"Theoretical Computer Science"}

READ

STRINGS { longest.label }

INTEGERS { number.label longest.label.width }

FUNCTION {initialize.longest.label}
{ "" 'longest.label :=
  #1 'number.label :=
  #0 'longest.label.width :=
}

FUNCTION {longest.label.pass}
{ number.label int.to.str$ 'label :=
  number.label #1 + 'number.label :=
  label width$ longest.label.width >
    { label 'longest.label :=
      label width$ 'longest.label.width :=
    }
    'skip$
  if$
}

EXECUTE {initialize.longest.label}

ITERATE {longest.label.pass}

FUNCTION {begin.bib}
{ preamble$ empty$
    'skip$
    { preamble$ write$ newline$ }
  if$
  "\begin{thebibliography}{"  longest.label  * "}" * write$ newline$
}

EXECUTE {begin.bib}

EXECUTE {init.state.consts}

ITERATE {call.type$}

FUNCTION {end.bib}
{ newline$
  "\end{thebibliography}" write$ newline$
}

EXECUTE {end.bib}

\end{document}


\title{Supplementary Information for: Surface roughening in nanoparticle catalysts}

\author{Cameron J. Owen$^{*,\dagger}$}
\affiliation{Department of Chemistry and Chemical Biology, Harvard University, Cambridge, Massachusetts 02138, United States}
\affiliation{John A. Paulson School of Engineering and Applied Sciences, Harvard University, Cambridge, Massachusetts 02138, United States}

\author{Nicholas Marcella$^{*}$}
\affiliation{Department of Chemistry, University of Illinois, Urbana, Illinois 61801, United States}

\author{Christopher R. O'Connor$^{*}$}
\affiliation{Rowland Institute at Harvard, Harvard University, Cambridge, Massachusetts 02142, United States}

\author{\\Taek-Seung Kim}
\affiliation{Rowland Institute, Harvard University, Cambridge, Massachusetts 02138, United States}

\author{Clare Yijia Xie}
\affiliation{John A. Paulson School of Engineering and Applied Sciences, Harvard University, Cambridge, Massachusetts 02138, United States}

\author{Ralph G. Nuzzo}
\affiliation{Department of Chemistry, University of Illinois, Urbana, Illinois 61801, United States}

\author{Anatoly I. Frenkel$^{\dagger}$}
\affiliation{Department of Materials Science and Chemical Engineering, Stony Brook University, Stony Brook, New York 11794, United States}
\affiliation{Chemistry Division, Brookhaven National Laboratory, Upton, New York 11973, United States}

\author{Christian Reece$^{*,\dagger}$}
\affiliation{Rowland Institute, Harvard University, Cambridge, Massachusetts 02138, United States}

\author{Boris Kozinsky$^{\dagger}$}
\affiliation{John A. Paulson School of Engineering and Applied Sciences, Harvard University, Cambridge, Massachusetts 02138, United States}
\affiliation{Robert Bosch LLC Research and Technology Center, Watertown, Massachusetts 02472, United States}

\def\thefootnote{$*$}\footnotetext{These authors contributed equally.}\def\thefootnote{\arabic{footnote}}

\def\thefootnote{$\dagger$}\footnotetext{Corresponding authors\\C.J.O., E-mail: \url{cowen@g.harvard.edu}\\A.I.F., E-mail: \url{anatoly.frenkel@stonybrook.edu}\\C.R., E-mail: \url{christianreece@fas.harvard.edu}\\B.K., E-mail: \url{bkoz@seas.harvard.edu}}\def\thefootnote{\arabic{footnote}}

\newcommand\bvec{\mathbf}
\newcommand{\mathsc}[1]{{\normalfont\textsc{#1}}}

\maketitle

\subsection*{Supplementary Notes}
\subsection*{Supplementary Note 1. Active-learning in FLARE for \textit{ab initio} data collection.}
Suppl. Table 1 contains a complete summary of the active learning trajectories used to build the \textit{ab initio } data set for MLFF training.

\begin{table*}[!htbp]
\centering
\resizebox{\textwidth}{!}{\begin{tabular}{|c|c|c|c|c|c|c|c|c|}
\multicolumn{1}{c}{\bf System}    & \multicolumn{1}{c}{\bf Temp. (K)} & \multicolumn{1}{c}{\bf $\Delta$ t (fs)}  &  \multicolumn{1}{c}{\bf $\sum\tau_{\textrm{sim}}$ (ns)} & \multicolumn{1}{c}{\bf   $\sum\tau_{\textrm{wall}}$ (hr)} & \multicolumn{1}{c}{\bf $\sum N_{\textrm{DFT}}$} & \multicolumn{1}{c}{\bf $N_{\textrm{atoms}}$} & \multicolumn{1}{c}{\bf $N_{\textrm{cpus}}$} \\
\hline
\multirow{2}{*}{CO/Pt(111)} & 500 & 0.2 & 0.10 & 85.6 & 59 & 96 & 96 \\ 
 & 1900 & 0.2 & 0.20 & 124.2 & 352 & 96 & 96 \\ 
\hline
\multirow{2}{*}{CO/Pt(100)} & 500 & 0.2 & 0.10 & 98.5 & 86 & 96 & 96 \\ 
 & 1900 & 0.2 & 0.20 & 176.8 & 385 & 96 & 96 \\ 
\hline
\multirow{2}{*}{Pt$_{43}$} & 1000 & 5.0 & 1.00 & 36.6 & 92 & 43 & 48 \\ 
 & 2100 & 5.0 & 0.39 & 130.0 & 60 & 43 & 48 \\ 
\hline
\multirow{2}{*}{CO$_{0.4 ML}$/Pt$_{43}$} & 1000 & 0.2 & 0.02 & 133.9 & 56 & 67 & 64 \\ 
 & 2100 & 0.2 & 0.005 & 117.1 & 50 & 67 & 64 \\ 
\hline
\multirow{2}{*}{CO$_{0.8 ML}$/Pt$_{43}$} & 1000 & 0.2 & 0.02 & 99.6 & 60 & 91 & 96 \\ 
 & 2100 & 0.2 & 0.0007 & 30.1 & 17 & 91 & 96 \\ 
\hline
\multirow{2}{*}{Pt$_{55}$} & 1000 & 5.0 & 0.33 & 40.7 & 107 & 55 & 64 \\ 
 & 2100 & 5.0 & 1.00 & 88.2 & 51 & 55 & 64 \\ 
\hline
\multirow{2}{*}{CO$_{0.35 ML}$/Pt$_{55}$} & 1000 & 0.2 & 0.027 & 112.9 & 58 & 91 & 96 \\ 
 & 2100 & 0.2 & 0.02 & 159.6 & 65 & 91 & 96 \\ 
\hline
\multirow{2}{*}{CO$_{0.71 ML}$/Pt$_{55}$} & 1000 & 0.2 & 0.026 & 63.8 & 41 & 115 & 128 \\ 
 & 2100 & 0.2 & 0.0001 & 8.0 & 6 & 115 & 128 \\ 
\hline
\multirow{2}{*}{Pt$_{79}$} & 1000 & 5.0 & 1.00 & 48.3 & 29 & 79 & 96 \\ 
 & 2100 & 5.0 & 0.49 & 164.3 & 50 & 79 & 96 \\ 
\hline
\multirow{2}{*}{CO$_{0.5 ML}$/Pt$_{79}$} & 1000 & 0.2 & 0.02 & 65.3 & 27 & 151 & 144 \\ 
 & 2100 & 0.2 & 0.0001 & 13.4 & 6 & 151 & 144 \\ 
\hline
Pt$_{135}$ & 1000 & 5.0 & 1.00 & 32.0 & 19 & 135 & 144 \\
\hline
CO$_{1.0 ML}$/Pt$_{135}$ & 1000 & 0.2 & 0.00001 & 8.0 & 7 & 207 & 288 \\
\hline
Pt$_{147}$ & 1000 & 5.0 & 1.00 & 18.7 & 11 & 147 & 160 \\
\hline
CO$_{0.5 ML}$/Pt$_{147}$ & 1000 & 0.2 & 0.00001 & 12.0 & 8 & 239 & 240 \\
\hline
CO$_{1.0 ML}$/Pt$_{147}$ & 1000 & 0.2 & 0.00001 & 14.2 & 7 & 331 & 336 \\
\hline
CO$_{gas}$ & 1900 & 0.2 & 0.10 & 10.1 & 55 & 96 & 96 \\
\hline
\end{tabular}}
\caption*{\textbf{Table 1: Summary of the parallel FLARE active learning trajectories.} The summary covers each of the CO, Pt, and CO/Pt bulk and nanoparticle systems, where the maximum simulation temperature (K), timestep (fs), total simulation time surveyed (ns), total hardware wall time employed (hr), total number of DFT calls during the simulation, number of atoms in the simulation cell, and number of cpu cores employed for each run.}
\end{table*}

\subsection*{Supplementary Note 2. MLFF locality test and length-scale interactions}
To verify the cutoff matrix used in our MLFF, we conducted the following locality test by displacing the an adsorbed CO molecule vertically and observe the distance dependance of the forces on the displaced CO and the total energy of CO/Pt slab system. Ultimately, our MLFF was developed with an angular resolution of $l_max=4$ and an interaction cutoff of $r_max=8$\AA{}. This cutoff was chosen due to the following single-point energy calculations demonstrating the length-scale of such interactions in the system, as shown in Suppl. Figs. 1-4. To construct such a demonstration, we performed single point energy calculations for a CO molecule both above a bare Pt(111) slab and above a full CO monolayer on Pt(111) at various separation distances. From the latter, we observe that for a CO molecule separating from the 1 ML adsorbate layer of CO,  the force and energy only plateau at $>7$ \AA{}. Hence, a maximum cutoff of 8 \AA{} was employed for the final model training.

\begin{table}[h!]
\centering
\caption{EXAFS Fitting parameters}
\begin{tabular}{@{}ccccccc@{}}
\toprule
\textbf{Temperature (K)} & \textbf{CN} & \textbf{$\Delta E_0$} & \textbf{$\sigma^2$} & \textbf{Theta} & \textbf{r} \\ \midrule
298  & 10.5 (3) & 7.8 (3) & 0.0068 (3) & 0.0001 (1) & 2.772 (8) \\
323  & 10.5 (3) & 7.8 (3) & 0.0074 (4) & 0.0002 (1) & 2.775 (8) \\
348  & 10.5 (3) & 7.8 (3) & 0.0077 (4) & 0.0002 (1) & 2.776 (8) \\
373 & 10.5 (3) & 7.8 (3) & 0.0079 (4) & 0.0001 (1) & 2.772 (8) \\
398 & 10.5 (3) & 7.8 (3) & 0.0081 (4) & 0.0002 (2) & 2.771 (9) \\
423 & 10.5 (3) & 7.8 (3) & 0.0082 (4) & 0.0003 (2) & 2.774 (9) \\
448 & 10.5 (3) & 7.8 (3) & 0.0085 (4) & 0.0003 (2) & 2.775 (8) \\
473 & 10.5 (3) & 7.8 (3) & 0.0089 (4) & 0.0003 (2) & 2.775 (8) \\
498 & 10.5 (3) & 7.8 (3) & 0.0090 (4) & 0.0003 (2) & 2.774 (9) \\
523 & 10.5 (3) & 7.8 (3) & 0.0094 (5) & 0.0002 (2) & 2.77 (1) \\
548 & 10.5 (3) & 7.8 (3) & 0.0095 (5) & 0.0003 (2) & 2.77 (1) \\
573 & 10.5 (3) & 7.8 (3) & 0.0096 (4) & 0.0004 (2) & 2.773 (9) \\ \bottomrule
\end{tabular}
\end{table}

\begin{figure*} 
\centering
\includegraphics[width=0.6\textwidth]{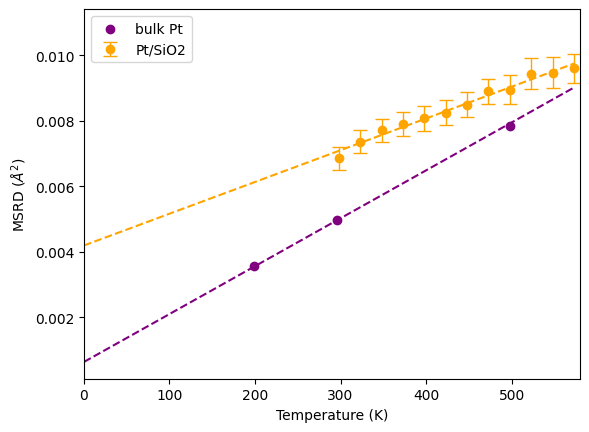}
\caption*{\textbf{Supplementary Figure 1:} MSRD of a bulk Pt reference sample and the supported Pt/SiO$_2$.}
\end{figure*}

\begin{figure*} 
\centering
\includegraphics[width=1.0\textwidth]{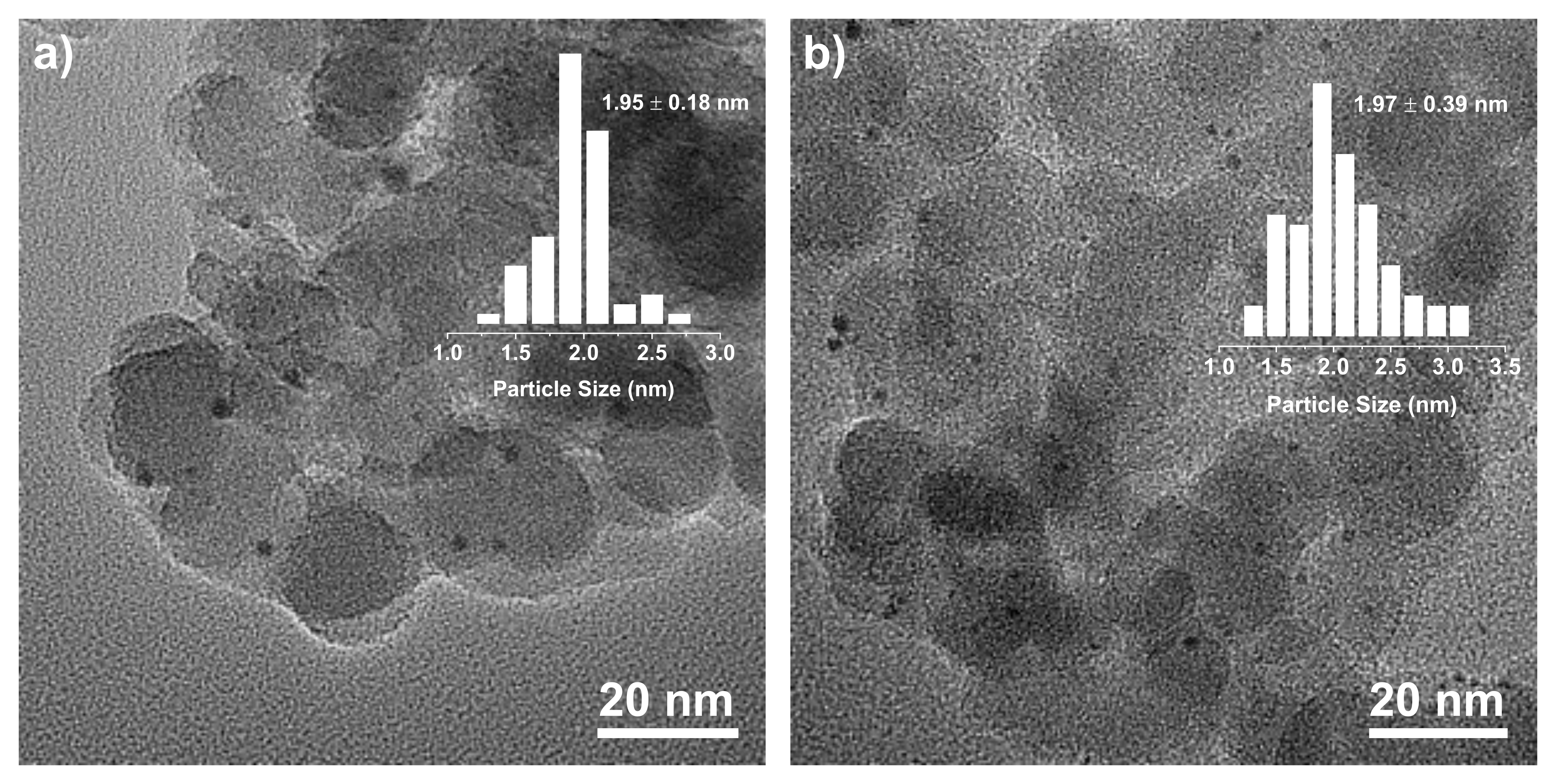}
\caption*{\textbf{Supplementary Figure 2:} Representative TEM images with particle size distributions (inset) of (a) fresh and (b) spent Pt/SiO$_2$ after the DRIFTS experiments demonstrate no significant change in particle size distribution.}
\end{figure*}

\begin{figure*} 
\centering
\includegraphics[width=0.6\textwidth]{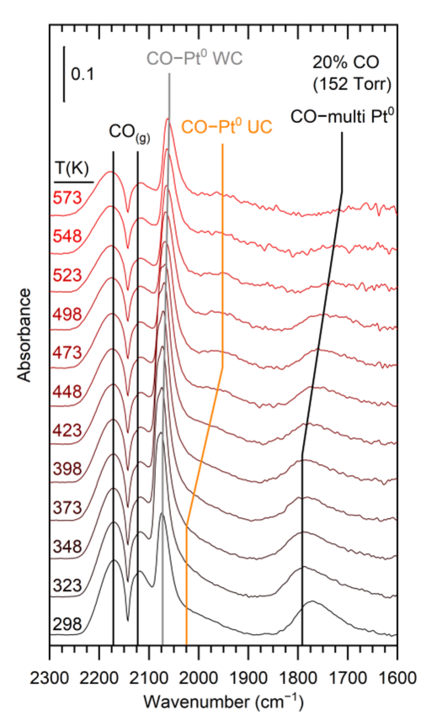}
\caption*{\textbf{Supplementary Figure 3:} Temperature dependent DRIFTS spectra of CO adsorbed on Pt/SiO$_2$ with the frequency positions of the three main features highlighted.}
\end{figure*}

\begin{figure*} 
\centering
\includegraphics[width=0.6\textwidth]{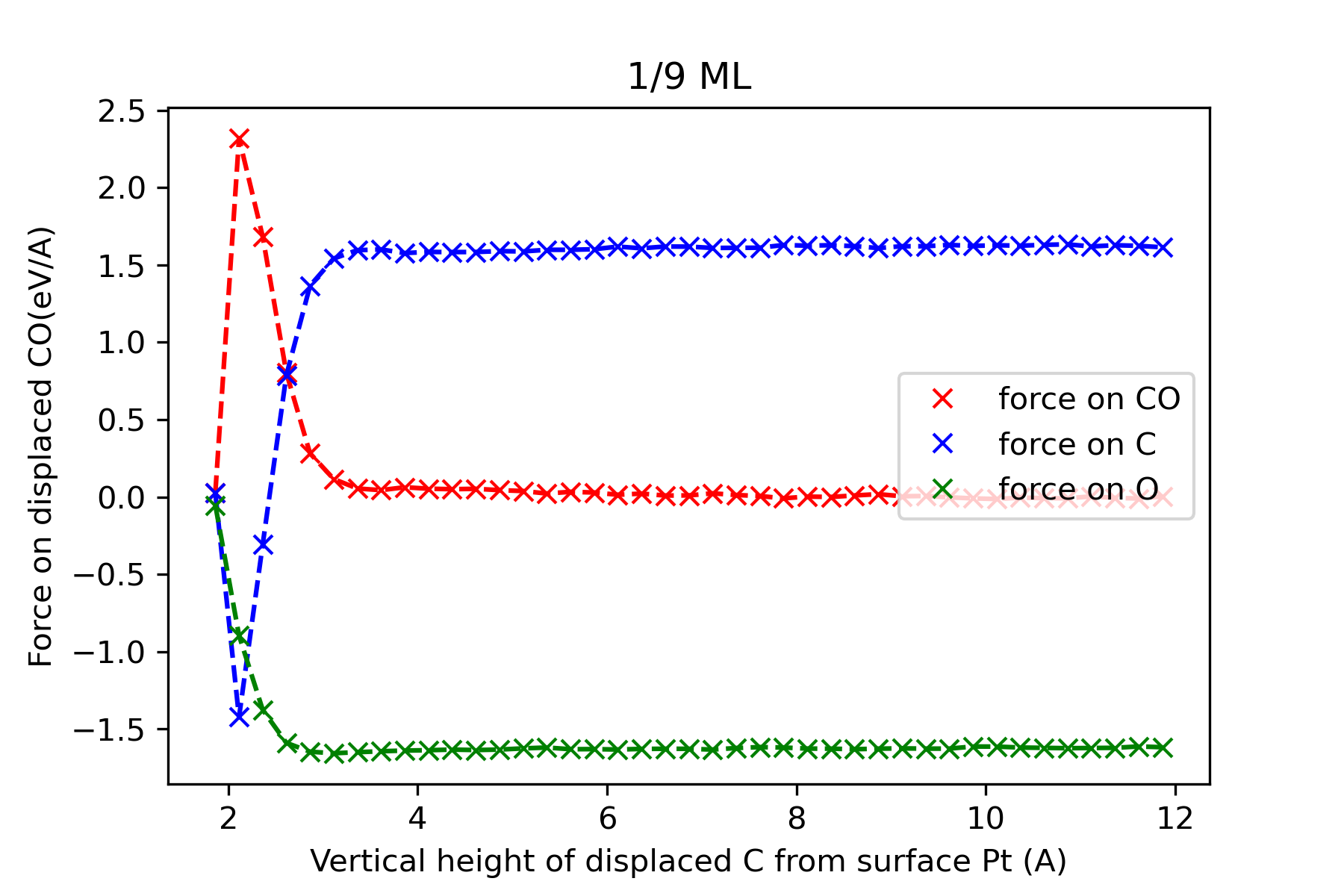}
\caption*{\textbf{Supplementary Figure 4:} Forces on displaced CO molecule and its component force on O and C atoms respectively as a function of veritical displacement for 1/9 ML coverage.}
\end{figure*}

\begin{figure*} 
\centering
\includegraphics[width=0.6\textwidth]{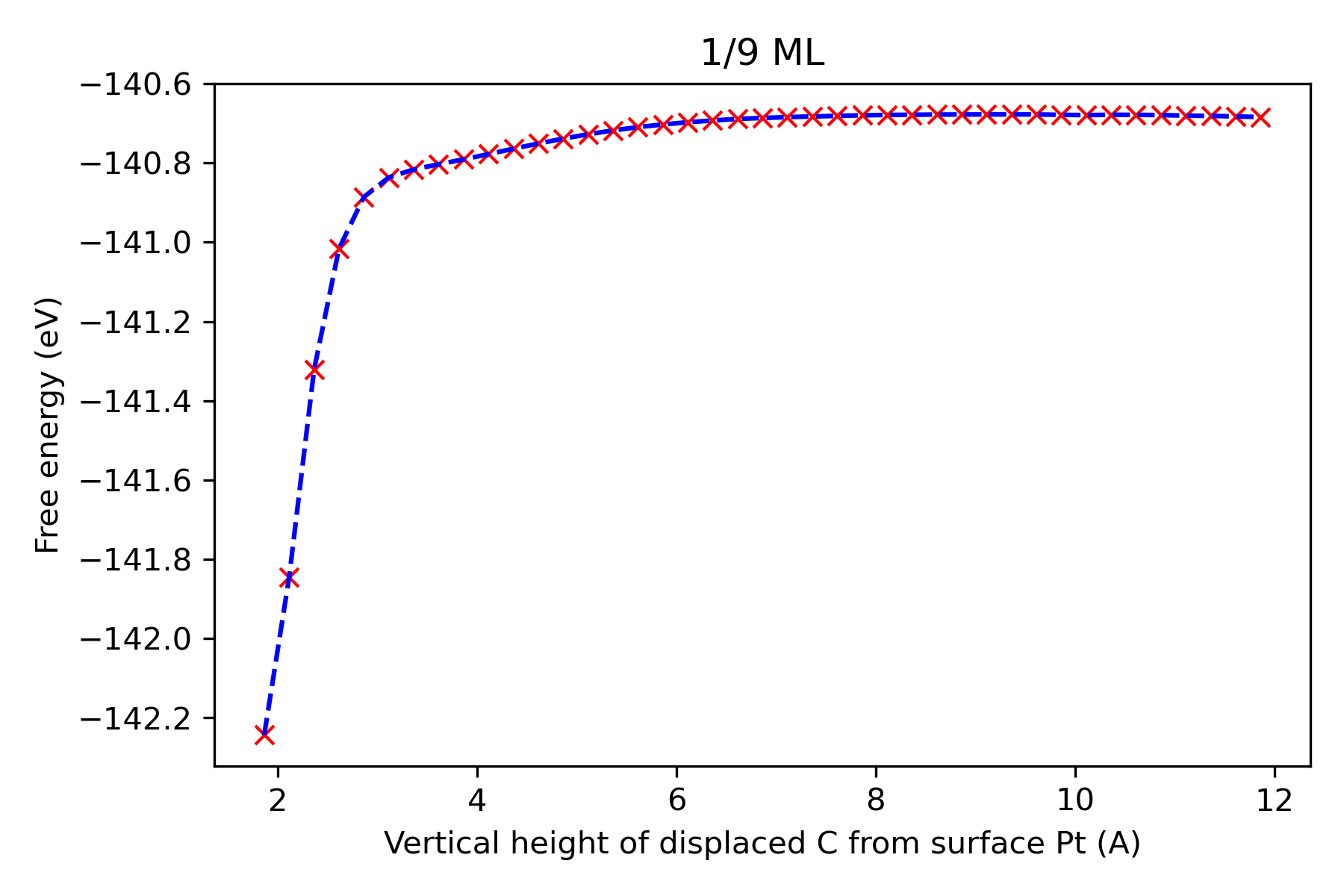}
\caption*{\textbf{Supplementary Figure 5:} Energies of CO adsorbed Pt slab with displaced CO as a function of vertical displacement for 1/9 ML coverage.}
\end{figure*}

\begin{figure*}
\centering
\includegraphics[width=0.6\textwidth]{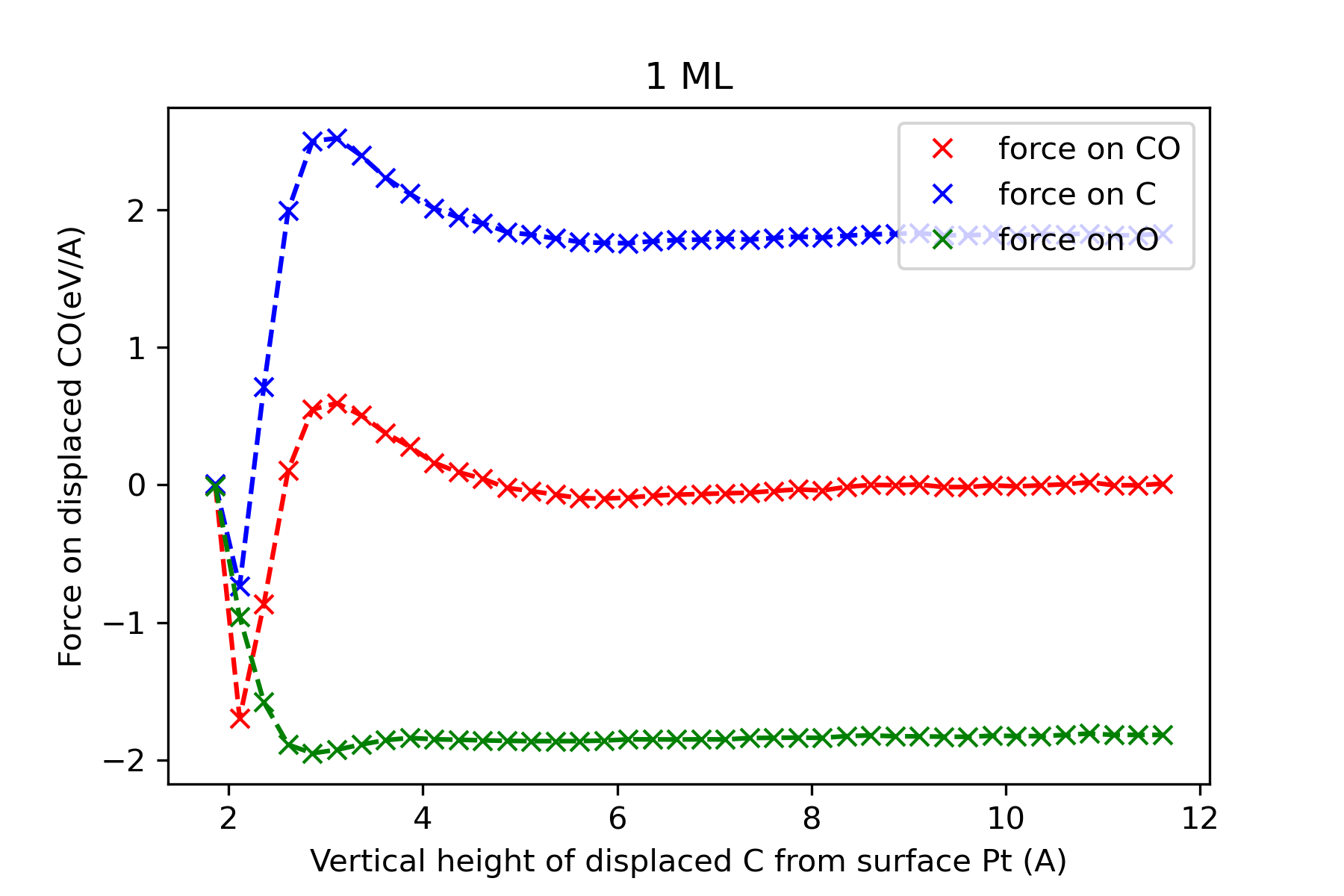}
\caption*{\textbf{Supplementary Figure 6:} Forces on displaced CO molecule and its component force on O and C atoms respectively as a function of vertical displacement for 1 ML coverage.}
\end{figure*}

\begin{figure*}
\centering
\includegraphics[width=0.6\textwidth]{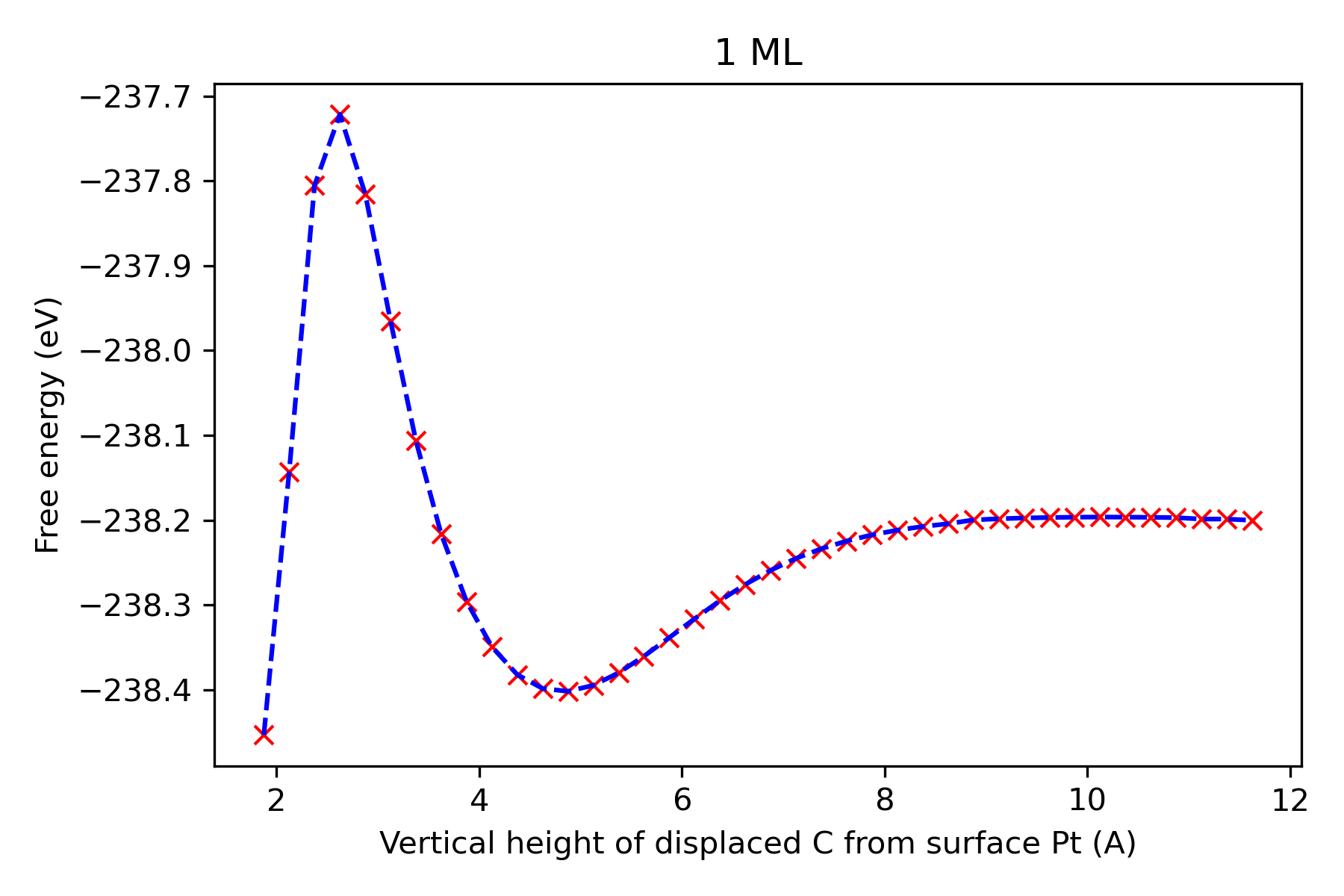}
\caption*{\textbf{Supplementary Figure 7:} Energies of CO adsorbed Pt slab with displaced CO as a function of vertical displacement for 1 ML coverage.}
\end{figure*}